\documentclass{jfm}
\usepackage{graphicx}
\usepackage{xcolor}
\usepackage{epstopdf, epsfig}

\shorttitle{Sedimentation of prolate spheroids in homogenous isotropic turbulence}
\shortauthor{Niazi Ardekani, Sardina, Brandt, Karp-Boss, Bearon and Variano}

\title{Sedimentation of elongated non-motile prolate spheroids in homogenous isotropic turbulence}

\author{M. Niazi Ardekani\aff{1}
  \corresp{\email{mehd@mech.kth.se}},
  G. Sardina\aff{1}, L. Brandt\aff{1}, L. Karp-Boss\aff{2}, R. N. Bearon\aff{3}
 \and E. A. Variano\aff{4}}

\affiliation{\aff{1}  Linn\'e Flow Centre and SeRC (Swedish e-Science Research Centre), KTH Mechanics, \\ S-100 44 Stockholm, Sweden
\aff{2} School of Marine Sciences, University of Maine, Orono, Maine 04469, USA
\aff{3} Department of Mathematical Sciences, University of Liverpool, Liverpool L69 7ZL, UK
\aff{4} Department of Civil and Environmental Engineering, University of California, \\ Berkeley, CA 94720, USA}
\begin{document}

\maketitle

\begin{abstract}
Phytoplankton are the foundation of aquatic food webs. Through photosynthesis, phytoplankton draw down $CO_2$ at magnitudes equivalent to forests and other terrestrial plants and convert it to organic material that is then consumed by other organisms of phytoplankton in higher trophic levels. Mechanisms that affect local concentrations and velocities are of primary significance to many encounter-based processes in the plankton including prey-predator interactions, fertilization and aggregate formation. We report results from simulations of sinking phytoplankton, considered as elongated spheroids, in homogenous isotropic turbulence to answer the question of whether trajectories and velocities of sinking phytoplankton are altered by turbulence. We show in particular that settling spheroids with physical characteristics similar to those of diatoms weakly cluster and preferentially sample regions of down-welling flow, corresponding to an increase of the mean settling speed with respect to the mean settling speed in quiescent fluid.  We explain how different parameters can affect the settling speed and what underlying mechanisms might be involved.  Interestingly, we observe that the increase in the aspect ratio of the prolate spheroids can affect the clustering and the average settling speed of particles by two mechanisms: first is the effect of aspect ratio on the rotation rate of the particles, which saturates faster than the second mechanism of increasing drag anisotropy.   

\end{abstract}

\begin{keywords}
Phytoplankton, prolate spheroids, tracers, settling particles, aspect ratio, homogenous isotropic turbulence   
\end{keywords}

\section{Introduction}

The behavior of suspended particles in turbulent flows is of considerable interest to a variety of processes \citep{Guazzelli2011} including dispersion of aerosols and collisions of water droplets in the atmosphere, dispersions of oil spills treated with dispersants, sediment transport and plankton ecology.  
The latter is perhaps the least familiar to fluid dynamicists but is of great importance because of the key roles plankton play in oceanic food webs, sustaining all the world's commercially important fisheries, and in global biogeochemical cycles, particularly the carbon cycle.  At the basis of all planktonic ecosystems are microscopic photosynthetic organisms, collectively called phytoplankton. These unicellular organisms reside in the upper layer of the ocean (and lakes), where both light and turbulence prevail, and are responsible for nearly $40\%$ of the global annual net primary productivity. They convert $CO_2$ into organic material that is consumed by other planktonic organisms and transferred to higher trophic levels.  A fraction of the organic material produced at the surface is transported to the deep ocean, via sinking particles, where it can become sequestered and remain there for thousands of years. Such sequestration is a critical piece in the global carbon budget.
Phytoplankton cells are generally denser than seawater and tend to sink \citep{Eppley1967}. Non-motile species of phytoplankton cannot control their vertical position in the water column and settling therefore determines vertical distributions and residence time of cells in the illuminated, upper layer of the ocean, ultimately affecting rates of photosynthesis and primary production. On much smaller scales, relative motion between the fluid and a sinking cell is sufficient to erode diffusive boundary layers and enhance fluxes of solutes to and from the cell \citep{Lazier1989,Karp1998}. Thus, mechanisms that affect sinking of phytoplankton hold significant implications to a variety of processes that govern carbon, nitrogen and energy flows in aquatic systems.  

To understand dynamics and residence time of phytoplankton in the upper mixed layer of oceans (and lakes) it is necessary to consider mechanisms that influence motions, trajectories and dispersion of individual cells. The mixed layer is defined as the layer between the ocean's surface and depth, typically ranging between $10$ to $200$ $m$, where turbulence, caused by wind stress, acts to mix physical and chemical properties. Superimposed on ambient turbulence are motions of phytoplankton due to swimming and sinking. Recent studies have advanced understanding of the effects of shear and turbulent flows on swimming behavior of phytoplankton \citep{Durham2011,Durham2013,cj,Barry2015,Gustavsson2016}, but interactions of turbulence with sinking cells remain largely unexplored.

Early observations that large phytoplankton cells, such as diatoms, thrive in turbulent waters and that stirring is required to maintain cells in suspension in culture flasks have led to the notion that turbulence increases residence time of phytoplankton in the mixed layer. This view, which became a paradigm in biological oceanography, is based on intuition rather than rigorous analysis.  Careful laboratory studies suggest the contrary- turbulence enhances sinking velocities of at least some species of phytoplankton, relative to their still-water velocities \citep{Ruiz2004} but mechanisms, magnitudes and consequences have not been fully characterized. Of particular interest, and the focus of this paper, is the manner in which turbulence biases particle trajectories, leading to altered sinking velocities. We particularly explain how different parameters can affect the settling speed and what underlying mechanisms might be involved.



Whether dissipation-scale oceanic turbulence can alter trajectories and lead to clustering of weakly inertial particles such as phytoplankton is still an open question. \cite{Nielsen1993} compiled data from experimental studies that examined a wide variety of suspensions, with both negatively and positively buoyant inclusions, showing the complex nature of turbulence-sinking (or rising) interactions. For weak to moderate turbulence, where ratios between vertical velocities of the flow and mean sinking or rising velocities in still water range between $0.3$ to $3$, both sinking of positively buoyant and rising of negatively buoyant particles are slowed by turbulence, presumably due to vortex trapping \citep{Tooby1977,Nielsen2007}. In strong turbulence, effects on heavy and light particles differ. While the sinking of heavy particles is accelerated, rising of light particles is further delayed by turbulence. The critical particle density separating delay from acceleration is not when the density of the particle matches that of the fluid \citep{Nielsen2007}. The observed increase in sinking velocities of heavy particles in strong turbulence has been attributed to the mechanism of fast tracking, where particles with sufficient inertia are preferentially swept out of vortices into regions of high strain and low vorticity and their trajectories tend to follow down-welling regions of the flow \citep{Murray1970, Maxey1987, Wang1993, Nielsen2007}.  Thus interactions of settling particles with turbulence can result in both 
preferential sampling of regions of specific velocity and clustering of particles. Mechanisms that affect local concentrations and velocities are of primary significance to many encounter-based processes in the plankton including prey-predator interactions, fertilization and aggregate formation \citep{Denny1993,Kiorboe2008}. 

\begin{figure}
\centering
\includegraphics[width=0.8\textwidth]{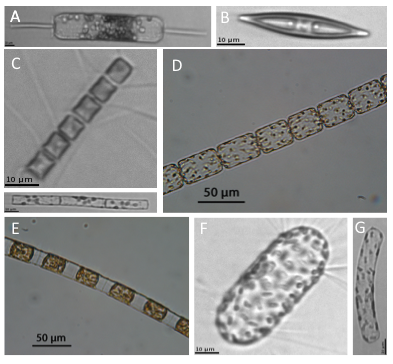}
\caption{\label{fig:diatoms}  
A mosaic of images of diatoms taken with a light microscope ($D$ \& $E$) or an Imaging Flow cytomter ($A$, $B$, $C$, $F$ \& $G$). A. \emph{Ditylum brightwellii} B. \emph{Navicula sp.} C. \emph{Chaetoceros sp.} (a chain forming diatom), D. \emph{Lauderia annulata} (a chain forming diatom), E. \emph{Lithodesmium undulatum} (a chain forming diatom), F. \emph{Chorethron sp.}, and G. \emph{Cylindrotheca sp.}} 
\end{figure}

Particle inertia is most commonly predicted by the Stokes number $St$ which defines the relative timescale it takes a particle to respond to a change in the flow compared to the timescale at which the flow itself changes.  The bulk of the literature is for $1 \leq St \leq 10$, while $St$ values for phytoplankton are $3$ to $4$ orders of magnitude smaller. $St$ by itself, however, may not be a good predictor of particle behavior \citep{Lucci2011,Fornari2016}. In addition to $St$, settling velocities, can play an important role, describing the particle behavior in turbulence flows.  

Conclusions from literature on cloud drops, oil droplets and aerosol do not transfer well to phytoplankton-like particles because of the difference in velocities, densities and size ratios \citep{Wang1993,Toschi2009,Gustavsson2014,Bec2014}, therefore there is a clear need to extend modelling efforts to the parameter space at which phytoplankton operate. A parameter space for modelling phytoplankton is given in more detail in appendix~\ref{app:Parameter_Space}. 

Another motivation for this study is to investigate the effect of shape on the dynamics of phytoplankton-like particles. So far only few studies have considered non-spherical particles in turbulent flows \citep{Ni2015,Byron2015,Parsa2014}. Phytoplankton exhibits a striking diversity of cell shapes and characteristics, which may influence their kinematics in important ways.  For example, \cite{Durham2011} showed that spherical phytoplankton cells that are bottom-heavy and swim upwards (gyrotactic swimmers) tend to get trapped and cluster at regions with high shear gradients in a vortical flow field and are more likely to be found in down-welling regions of turbulent flows \citep{Durham2013}. This tendency to cluster, however, appears to be sensitive to the shape of cells \citep{cj,Gustavsson2016}. We are not aware of similar studies on the effect of the cell shape for settling, non-motile phytoplankton.

 Here we use analytical and direct numerical simulations to examine the behavior of chain-forming diatoms in a turbulent flow. We focus on diatoms because this group is both ecologically important and promises diverse kinematics due to their relatively large size and diverse morphologies (e.g. Figure~\ref{fig:diatoms}). Regarding their ecological importance, diatoms often dominate primary production in seasonal blooms and are considered key players in the biological pump that removes carbon from the surface layer that is in contact with the atmosphere to the deep ocean. Regarding their size and morphology, although unicellular, many species form colonies in the form of long chains that can reach up to a few $mm$ in size. Furthermore, diatoms have biomineralized siliceous cell walls that contribute a significant, and variable, excess density of cells. The behavior of these cells in a simple, steady shear flow has been studied in the lab \citep{Karp1998}. While diatoms morphologies are complex and may include spines and other projections, as shown in figure~\ref{fig:diatoms}, \cite{Nguyen2011} showed that in a simple shear flow the motion of a diatom with spines or projections could be well predicted from theory of the motion of spheroids in a simple shear flow, providing that the cell is described by the smallest inscribing spheroid that encompasses both the cell and its spines.  We therefore model diatom chains as prolate spheroids. 
 
We begin by introducing the kinematic and dynamic models, used in this study to describe the particle motion in \S\ref{sec:Governing}. We compare the models and validate our numerical code in \S\ref{sec:Validation}. Next, we examine whether turbulence affects sinking velocities and clustering of weakly inertial prolate spheroids and how it may vary as a function of different particle parameters (\S\ref{sec:Results}). Alteration of sinking velocities is explained further in this section, considering a simple shear flow. The main conclusions and final remarks are drawn in \S\ref{sec:Final_remarks}. 
 
\section{Governing equations and numerical method}
\label{sec:Governing}

\subsection{Fluid phase}

The incompressible turbulent velocity field, $\textbf{u}$, obeys the Navier-Stokes and continuity equations:
\begin{eqnarray}
\label{eq:NS}  
\frac{\partial \textbf{u}}{\partial t} + \textbf{u} \cdot \nabla \textbf{u}  &=& -\nabla p + \frac{1}{Re} \nabla^2 \textbf{u} +  \textbf{f} \, , \\ 
\nabla \cdot \textbf{u} &=& \, 0 \, .
\end{eqnarray}
Here $t$ denotes the time, $p$ the pressure, $Re$ the Reynolds number and  $\textbf{f}$ the random forcing field necessary to maintain the turbulent velocity in a statistically steady state. The equations are discretized in a cubic domain with periodic conditions at the boundaries. We employ Direct Numerical Simulation (DNS) to solve all the relevant flow scales without any artificial model at the smallest scales. Since we evolve the equations in a three-periodic domain, it is natural to do so in Fourier space with a pseudo-spectral method. In this numerical scheme the nonlinear terms are evaluated in physical space using the classic $2/3$ rule to minimize aliasing error. Time integration is performed with a third-order low-storage Runge-Kutta method where the diffusive terms are analytically calculated while an Adam-Bashforth-like approximation is employed for the nonlinear terms \citep{rogallo}.
The  stochastic forcing is evaluated in Fourier space and acts isotropically on the first shell of wave vectors. The forcing amplitude is constant in time and the field is delta-correlated in time and uniformly distributed in phase and directions \citep{vinmen}. 
We employed a resolution of $128^3$ grid points with a Taylor Reynolds number $Re_\lambda=100$ and $64^3$ grid points with a Taylor Reynolds number $Re_\lambda=60$. The turbulence is characterized by the Taylor Reynolds number $Re_\lambda \equiv u^{\prime} \lambda / \nu$, where $u^{\prime}$ is the root-mean-square of velocity fluctuations, $\nu$ is the kinematic viscosity and $\lambda$ is the Taylor microscale, $\lambda \equiv \sqrt{ \epsilon / 15 \nu u^{\prime \, 2} }$, with $\epsilon$ the kinetic energy dissipation rate. The resolution was chosen so that the velocity gradients are well resolved since they drive the the equation of motion of the particles. 
Particle statistics are computed from independent samples separated by $1.6$ turbulent Kolmogorov time scales, $\tau_f \equiv \sqrt{\nu / \epsilon}$, corresponding to about $300$ configurations per particle. 
The numerical code has been already employed to study gyrotactic micro-organisms in homogeneous turbulence \citep{cj} and rain droplet evaporation in clouds \citep{prlcloud}.

\subsection{Solution of the particle motion}

Two different approaches are adopted to evaluate the dynamics of sedimenting spheroids in a Lagrangian framework. Diatoms are typically smaller than the characteristic Kolmogorov scale in the upper mixed layer of the ocean (see appendix~\ref{app:Parameter_Space}) and so are the particles in our model. The flow properties at the particle position are evaluated via linear interpolation from the neighbouring grid points. 

\subsubsection{Method I - Kinematic model}

Given the parameters relevant to plankton dynamics, the first approach assumes that the cells behave like passive tracers with a correction given by their Stokes settling velocity. The particle translational motion obeys

\begin{equation}
\frac{ \mathrm{d} \textbf{x}}{\mathrm{d} t} = \textbf{u} |_\mathrm{x} + \textbf{v}_s \,(\textbf{p}) \, ,
\label{eq:IntegTracers}  
\end{equation}
\begin{equation}
\textbf{v}_s \, (\textbf{p}) = v_s^{min} \hat{\textbf{e}}_g + \left (v_s^{max}-v_s^{min}\right ) \left(\hat{\textbf{e}}_g \cdot \textbf{p} \right) \, \textbf{p} \, ,
\label{eq:VelocityTracers}  
\end{equation}
with $\textbf{x}$ denoting spheroid position, $\textbf{u} |_\mathrm{x}$ the fluid velocity at spheroid position,  $\textbf{p}$ the spheroid orientation vector, $\textbf{v}_s$ the Stokes settling velocity and $\hat{\textbf{e}}_g$ a unit vector in the direction of gravity. 
$v_s^{min}$ and $v_s^{max}$ are the minimum and maximum settling speeds in quiescent flow, corresponding to the particles falling with their broad side perpendicular or parallel to the gravity direction. These velocities can be written as

\begin{equation}
v_s^{max} = \frac{\left( \rho_p - \rho_f \right) g l^2}{\mu_f} \gamma_0 \left( \mathcal{AR} \right) \,\,\,\, ; \,\,\,\, v_s^{min} = \frac{\left( \rho_p - \rho_f \right) g l^2}{\mu_f} \gamma_1 \left( \mathcal{AR} \right)  \, ,
\label{eq:VSmaxandVSmin}  
\end{equation}
with $g$ the gravitational acceleration, $l$ the length of the spheroid's major axis, $\mu_f$ the dynamic viscosity of the fluid, and $\rho_p$ and $\rho_f$ the densities of the particle and the fluid respectively. $\gamma_0$ and $\gamma_1$ are functions of the aspect ratio $\mathcal{AR}$, whose full expression can be found in  \cite{Dahlkild2011} among others. $\mathcal{AR}$ is defined as the ratio of polar over equatorial radii of the spheroid.

The spheroid rotation follows the inertialess Jeffery equation

\begin{equation}
\frac{ \mathrm{d} \textbf{p}}{\mathrm{d} t} = \frac{1}{2} \, \pmb{\omega} |_\mathrm{x} \times \textbf{p} + \alpha \left[\textbf{I} - \textbf{p}\textbf{p} \right] \cdot \textbf{E}|_\mathrm{x} \cdot \textbf{p}  \, ,
\label{eq:RotationTracers}  
\end{equation}
with $\pmb{\omega} |_\mathrm{x}$ the flow vorticity at the particle's position, $\textbf{I}$ the second order identity tensor, $\textbf{E}|_\mathrm{x}$ the symmetric part of the velocity deformation tensor and $\alpha$ a function of the spheroid aspect ratio $\mathcal{AR}$, defined as

\begin{equation}
\alpha = \frac{\mathcal{AR}^2 - 1}{\mathcal{AR}^2 + 1} \, .
\label{eq:AspectRatio}  
\end{equation}

The same Runge-Kutta scheme used for the carrier phase is used to perform the time integration for equation \ref{eq:RotationTracers}. This  is evaluated via a formulation based on quaternions. We evolve a total of $200000$ spheroids for each simulation.

\subsubsection{Method II - Dynamic model}


In the second approach, the spheroids are assumed to be inertial. They are assumed to obey the classic Maxey-Riley equation with an orientation correction for the Stokes drag \citep{Gallily1979,brenner,zhao}, without the Basset contribution:

\begin{equation}
\frac{ \mathrm{d} \textbf{v}}{\mathrm{d} t} = \textbf{A}^t \textbf{K}^\prime \textbf{A} \frac{\textbf{u} - \textbf{v}}{\tau_p} + \frac{\rho_f}{\rho_p} \frac{ \mathrm{D} \textbf{u}}{\mathrm{D} t}  + \frac{1}{2} \frac{\rho_f}{\rho_p} \left[ \frac{ \mathrm{D} \textbf{u}}{\mathrm{D} t} - \frac{\mathrm{d} \textbf{v}}{\mathrm{d} t} \right] + \left( 1 -  \frac{\rho_f}{\rho_p} \right) \textbf{g}  \, ,
\label{eq:IntegForces1}  
\end{equation}
\begin{equation}
\frac{ \mathrm{d} \textbf{x}}{\mathrm{d} t} = \textbf{v} \, ,
\label{eq:IntegTracers2}  
\end{equation}
where the terms on the right-hand-side of equation \ref{eq:IntegForces1} represent
from left to right: the Stokes drag, the pressure gradient, added mass and lastly gravity (also accounting for the buoyancy force). The Basset history term  is neglected in the present work due to its small contribution to the particle acceleration in the range of Stokes number and density ratio studied here (table~\ref{table:Sediment}) \citep[see][]{Olivieri2013,Olivieri2014,Daitche2015}. In the expression above, $\textbf{v}$ is the particle velocity, $\textbf{A}$ is the transformation tensor between the fixed and the local spheroid reference frames, $*^t$ denotes the transpose operator, 
and
$\tau_p$  the relaxation time for prolate spheroids, defined by
\begin{equation}
\tau_p =  \frac{2  \rho_p a^2}{9 \mu_f} \, \, \frac{ \mathcal{AR} \ln ( \mathcal{AR} + \sqrt{\mathcal{AR}^2 - 1} ) }{\sqrt{\mathcal{AR}^2 - 1}} \, ,
\label{eq:taup}  
\end{equation}
with $a$ the semi-minor axis of the prolate spheroid. $D */Dt$ is the fluid material derivative, $\textbf{g}$ the gravity vector.
$\textbf{K}^\prime$ is the resistance second order tensor in the spheroid reference frame; it is a diagonal tensor of components, in which $z$ denotes the polar axis.
\begin{equation}
K^\prime_{xx} = K^\prime_{yy} = \frac{(16/6) (\mathcal{AR}^2 - 1) \ln ( \mathcal{AR} + \sqrt{\mathcal{AR}^2 - 1} )}{(2 \mathcal{AR}^2 - 3) \ln ( \mathcal{AR} + \sqrt{\mathcal{AR}^2 - 1} ) + \mathcal{AR} \sqrt{\mathcal{AR}^2 - 1}} \, ,
\label{eq:kxx}  
\end{equation}

\begin{equation}
K^\prime_{zz}  = \frac{(8/6) (\mathcal{AR}^2 - 1) \ln ( \mathcal{AR} + \sqrt{\mathcal{AR}^2 - 1} )}{(2 \mathcal{AR}^2 - 1) \ln ( \mathcal{AR} + \sqrt{\mathcal{AR}^2 - 1} ) - \mathcal{AR} \sqrt{\mathcal{AR}^2 - 1}} \, .
\label{eq:kzz}  
\end{equation}

The modified Maxey-Riley equation (\ref{eq:IntegForces1}) is re-written in terms of relative velocity $\textbf{w}=\textbf{v}-\textbf{u}$ for a more accurate numerical solution,

\begin{equation}
\left( 1 +  \frac{1}{2} \frac{\rho_f}{\rho_p} \right) \frac{ \mathrm{d} \textbf{w}}{\mathrm{d} t} = -\textbf{A}^t \textbf{K}^\prime \textbf{A} \frac{\textbf{w}}{\tau_p} + \left( \frac{\rho_f}{\rho_p} -1 \right) \frac{ \mathrm{d} \textbf{u}}{\mathrm{d} t}  - \frac{3}{2} \frac{\rho_f}{\rho_p} \textbf{w} \cdot  \nabla \textbf{u} + \left( 1 -  \frac{\rho_f}{\rho_p} \right) \textbf{g} \, ,
\label{eq:IntegForces2}  
\end{equation}
and is advanced in time with an implicit predictor-corrector scheme to avoid numerical instabilities associated with the small values of the spheroid relaxation time.

The spheroid rotation is governed by the following equation
\begin{equation}
\frac{ \mathrm{d} \left(\textbf{I}^\prime \cdot \pmb{\omega}^\prime \right) }{\mathrm{d} t}  \, + \, \pmb{\omega}^\prime \times \left(\textbf{I}^\prime \cdot\pmb{\omega}^\prime \right) =  \textbf{N}^\prime \, ,
\label{eq:RotationForces}  
\end{equation}
with $\textbf{I}^\prime$ the spheroid moment of inertia tensor, $\pmb{\omega}^\prime$ the spheroid angular velocity and $\textbf{N}^\prime$ the rotation torque, where the superscript $*^\prime$ denotes quantities in the spheroid reference frame.
$\textbf{N}^\prime$ is given by 

\begin{eqnarray}
N^\prime_{x} &=& C_x \left [ \left ( 1-\mathcal{AR}^2\right)E^\prime_{yz} +\left ( 1+\mathcal{AR}^2\right)\left ( \Omega^\prime_x-\omega^\prime_x\right) \right] \,\, ; \,\, C_x=\frac{16\pi\mu_fa^3\mathcal{AR}}{3 \left (\beta_0+\mathcal{AR}^2\gamma_0  \right)} \, ,	\\
N^\prime_{y} &=& C_y \left [ \left ( 1-\mathcal{AR}^2\right)E^\prime_{xz} +\left ( 1+\mathcal{AR}^2\right)\left ( \Omega^\prime_y-\omega^\prime_y\right) \right] \,\, ; \,\,  C_y=\frac{16\pi\mu_fa^3\mathcal{AR}}{3 \left (\beta_0+\mathcal{AR}^2\gamma_0  \right)}  \, , \\
N^\prime_{z} &=& C_z \left ( \Omega^\prime_z-\omega^\prime_z\right) \,\, ; \,\, C_z=\frac{32\pi\mu_fa^3\mathcal{AR}}{3 \left (\alpha_0+\beta_0\right )} \, ,
\label{eq:n}  
\end{eqnarray}
with ${\bf \Omega}^\prime$ is the fluid rate of rotation tensor, and $\alpha_0$, $\beta_0$ and $\gamma_0$ are coefficient functions of the aspect ratio whose expression can be found in \cite{jeff}.

\section{Validation and comparison between the two models}
\label{sec:Validation}

\begin{figure}
\centering
\includegraphics[width=0.495\linewidth]{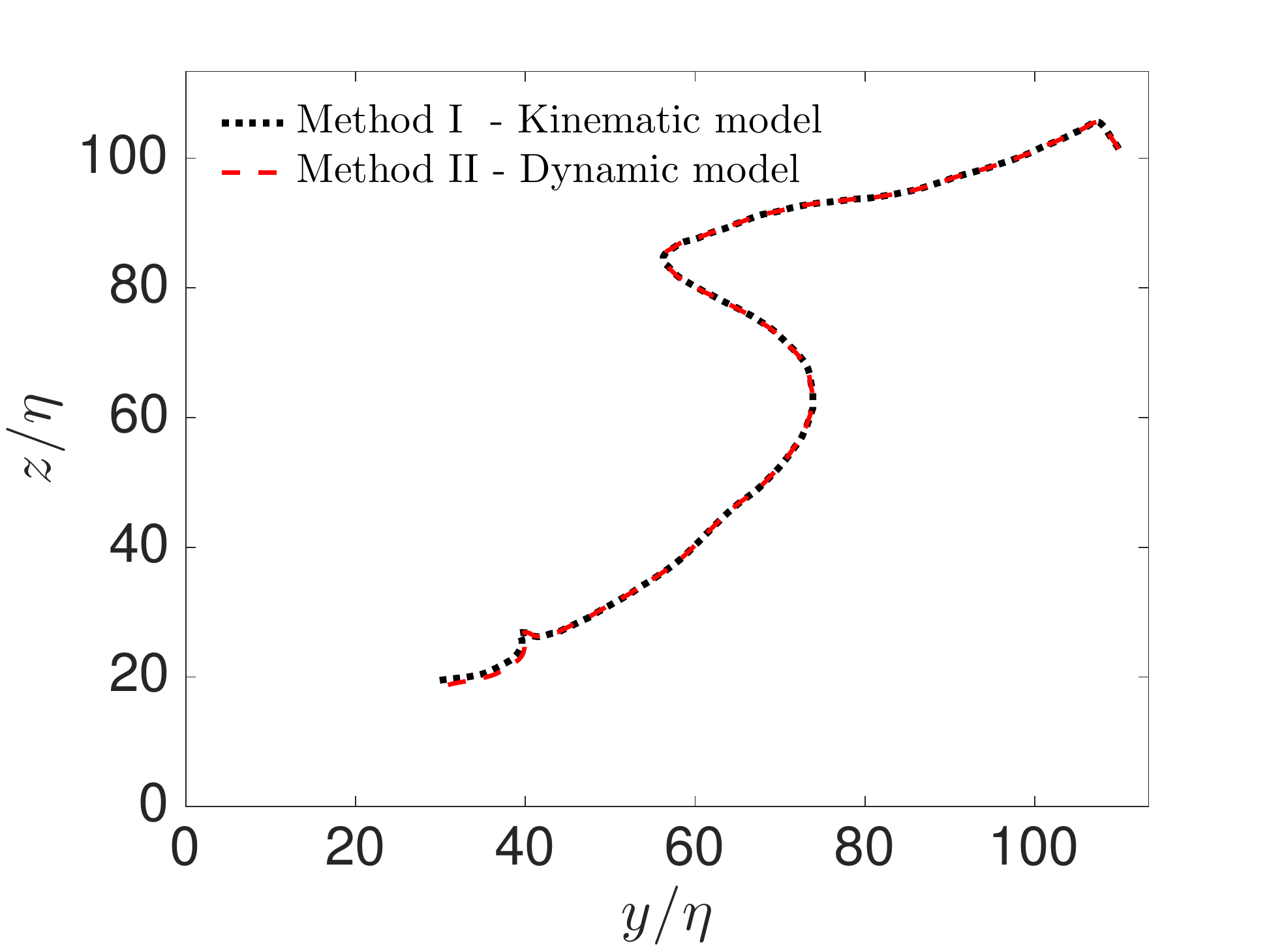}
\includegraphics[width=0.495\linewidth]{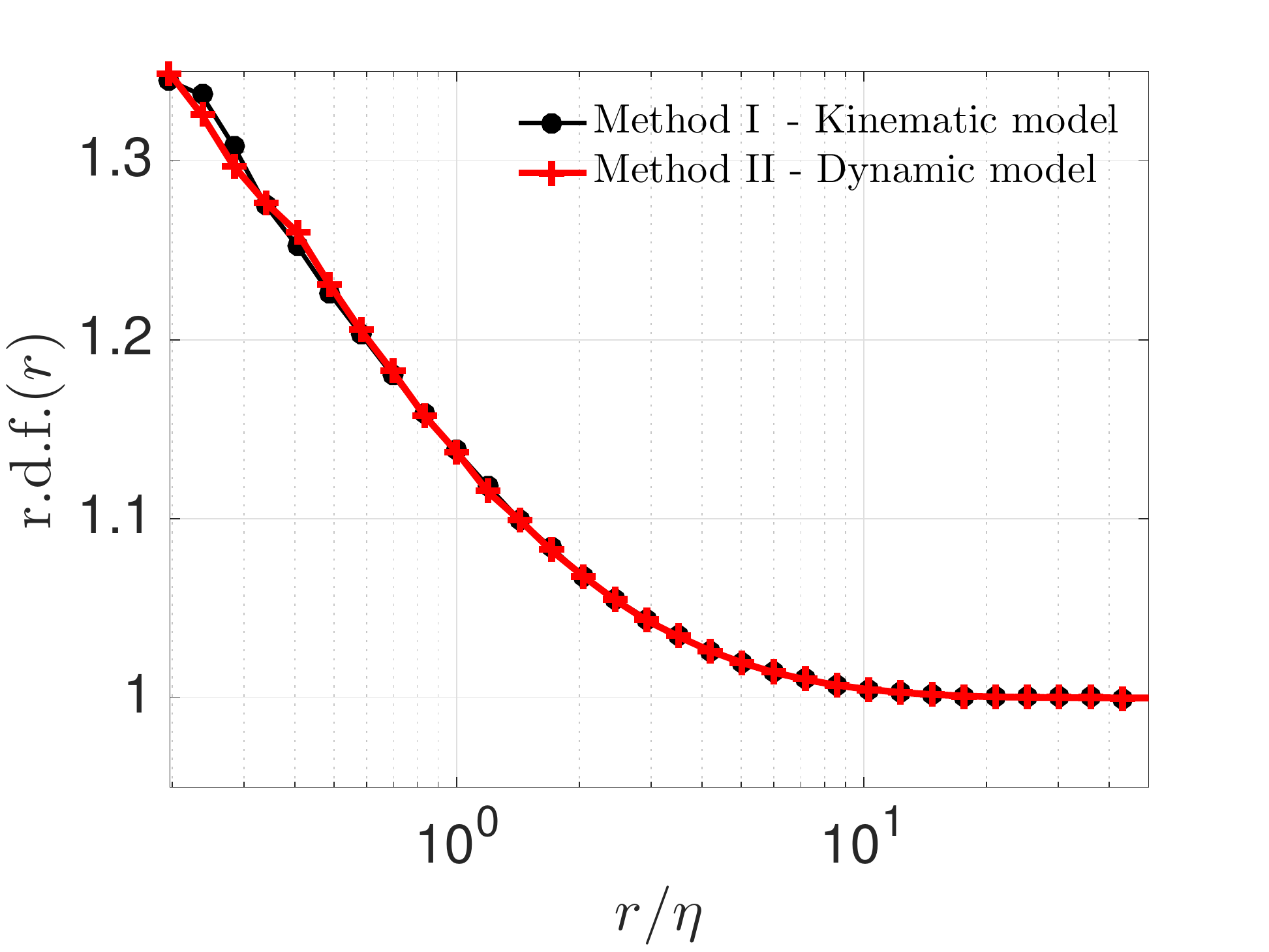}
\put(-395,120){$(a)$}
\put(-197,120){$(b)$}
\caption{\label{fig:valid1} 
Comparison between kinematic (non-inertial; black line) and dynamic (inertial; red line) models: $(a)$ Example trajectories of the two particles, using the two models from $t=0$ to $t=40\tau_\eta$ and $(b)$ radial distribution function (r.d.f) of particle position. $Re_\lambda = 60$ ; $\mathcal{AR}=5$ ; $\rho_p / \rho_f = 1.3$ ; $v^{iso}_s \approx 2.3 u_\eta$. $v^{iso}_s$ is the average settling speed for spheroids in a quiescent flow and the r.d.f measures the probability to find a particle pair at a given radial distance, normalized by the values of a uniform distribution. (see \S~\ref{sec:Results} for the formulation of r.d.f. and $v^{iso}_s$).} 
\end{figure}

\begin{figure}
\centering
\includegraphics[width=0.495\linewidth]{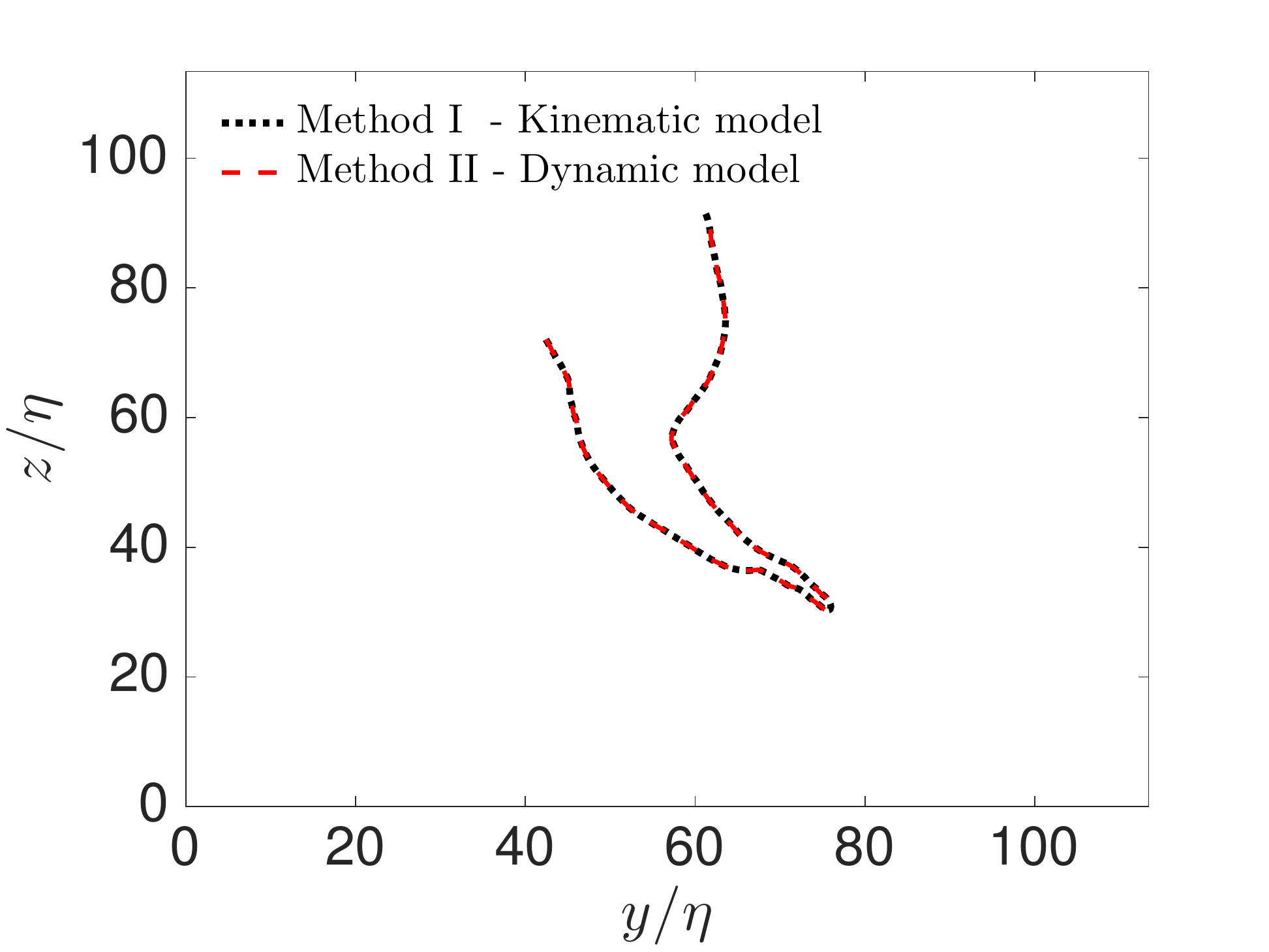}
\includegraphics[width=0.495\linewidth]{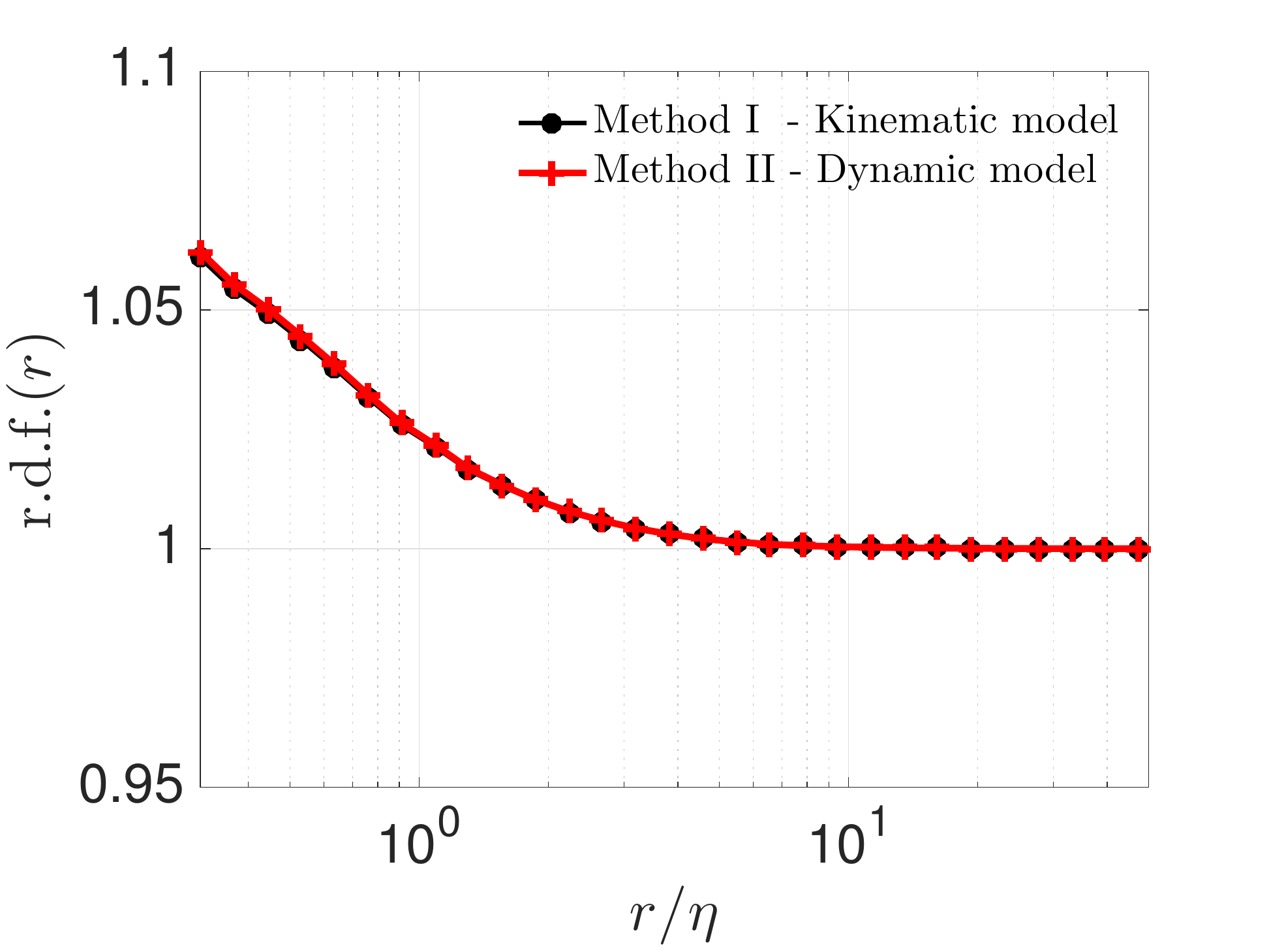}
\put(-395,120){$(a)$}
\put(-197,120){$(b)$}
\caption{\label{fig:valid2} 
Comparison between kinematic (non-inertial; black line) and dynamic (inertial; red line) models: $(a)$ Example trajectories of the two particles, using the two models from $t=0$ to $t=40\tau_\eta$ and $(b)$ radial distribution function (r.d.f) of particle position.
$Re_\lambda = 60$ ; $\mathcal{AR}=5$ ; $\rho_p / \rho_f = 1.05$ ; $v^{iso}_s \approx 1.1 u_\eta$. $v^{iso}_s$ is the average settling speed for spheroids in a quiescent flow and the r.d.f measures the probability to find a particle pair at a given radial distance, normalized by the values of a uniform distribution. (see \S~\ref{sec:Results} for the formulation of r.d.f. and $v^{iso}_s$).} 
\end{figure}

We first compare the two models to assess whether the simpler, inertialess model can be used for simulations aiming to study the settling of plankton in turbulent flow (see table~\ref{tab:param}). For each method, we perform two simulations at $Re_\lambda$ around $60$ and spheroid aspect ratio of $\mathcal{AR}=5$ . The two cases differ in the choice of density ratio ($1.3$ in figure~\ref{fig:valid1} and $1.05$ in figure~\ref{fig:valid2}) and average settling velocity ($2.3 u_\eta$ in figure~\ref{fig:valid1} and $1.1 u_\eta$ in figure~\ref{fig:valid2}).

We consider both particle trajectories and statistical properties: the two different methods provide the same result for both the single spheroid trajectories (panels $(a)$ of figures \ref{fig:valid1} and \ref{fig:valid2}) and the global spheroid two-point distribution function (panels $(b)$ of figures \ref{fig:valid1} and \ref{fig:valid2}).
As seen in these figures, in the limit of  small density ratio and relaxation time the two models provide the same dynamics. Therefore, we have used the non-inertial tracer (kinematic) model for the simulations at higher Reynolds number that will be described in the next sections.

\section{Results} 
\label{sec:Results} 

\begin{table}
\centering
\begin{tabular}{l | *{6}{c}}
$\mathcal{AR}$ & $1$ & $2$ & $3$ & $5$ & $10$ & $20$ \\
\hline
 $v_s^{max}$ & $3.03$ & $3.17$ & $3.11$ & $2.90$ & $2.46$ & $1.97$ \\ 
 $v_s^{min}$ & $3.03$ & $2.77$ & $2.53$ & $2.18$ & $1.71$ & $1.29$\\ 
 $v_s^{max} / v_s^{min}$ & $1$ & $1.14$ & $1.23$ & $1.33$ & $1.44$ & $1.53$ \\
 $v^{iso}_s$ & $3.03$ & $2.90$ & $2.72$ & $2.42$ & $1.96$ & $1.52$ \\ 
\end{tabular} 
\caption{\label{table:Sediment} Maximum, minimum and average of settling velocities in quiescent flow, scaled with the Kolmogorov velocity for the investigated cases with different aspect ratios. $\rho_p / \rho_f = 1.05$ and $D_{eq} = \eta/6$.} 
\end{table}

\begin{figure}
\centering
\includegraphics[width=0.495\linewidth]{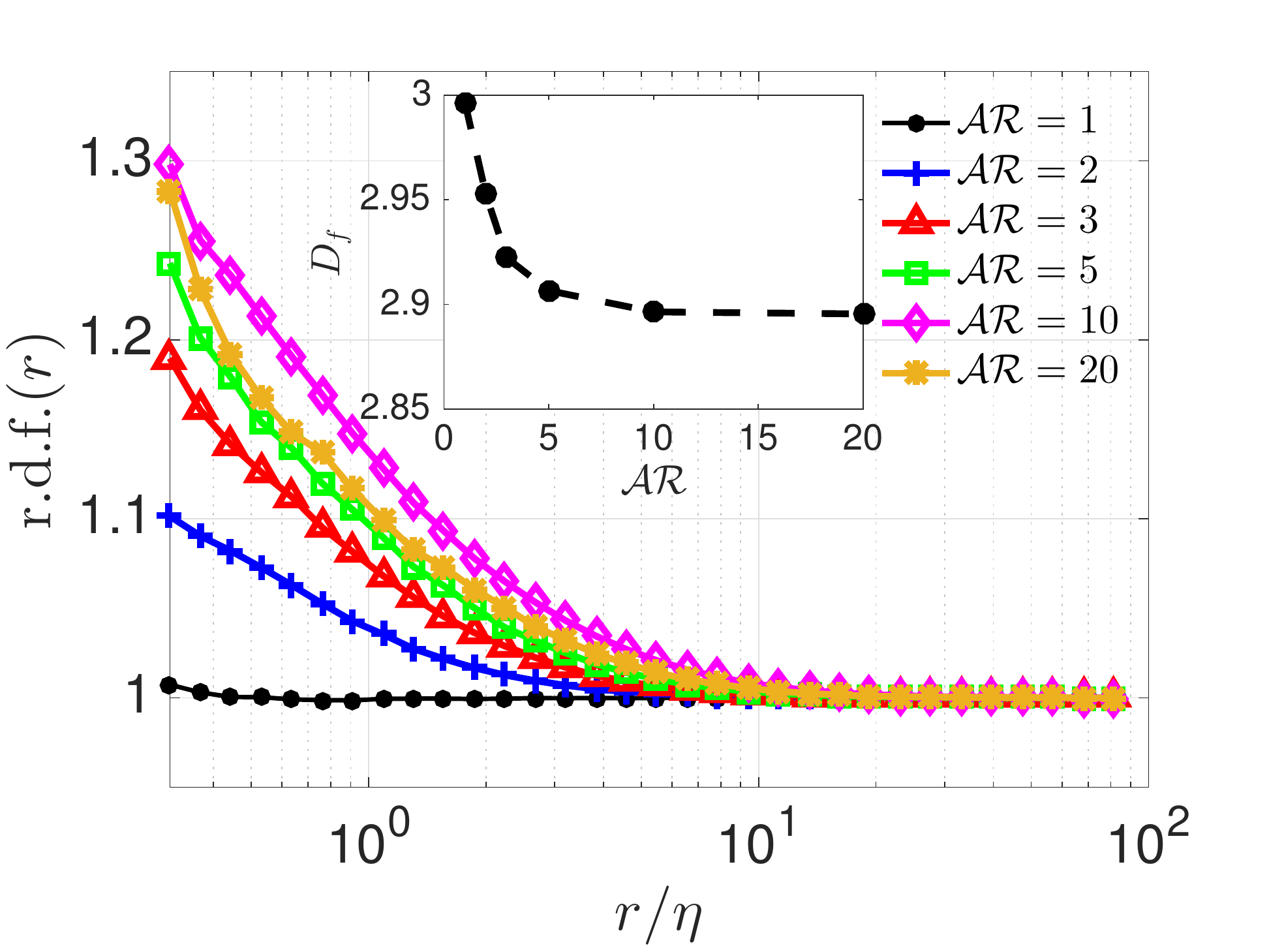}
\includegraphics[width=0.495\linewidth]{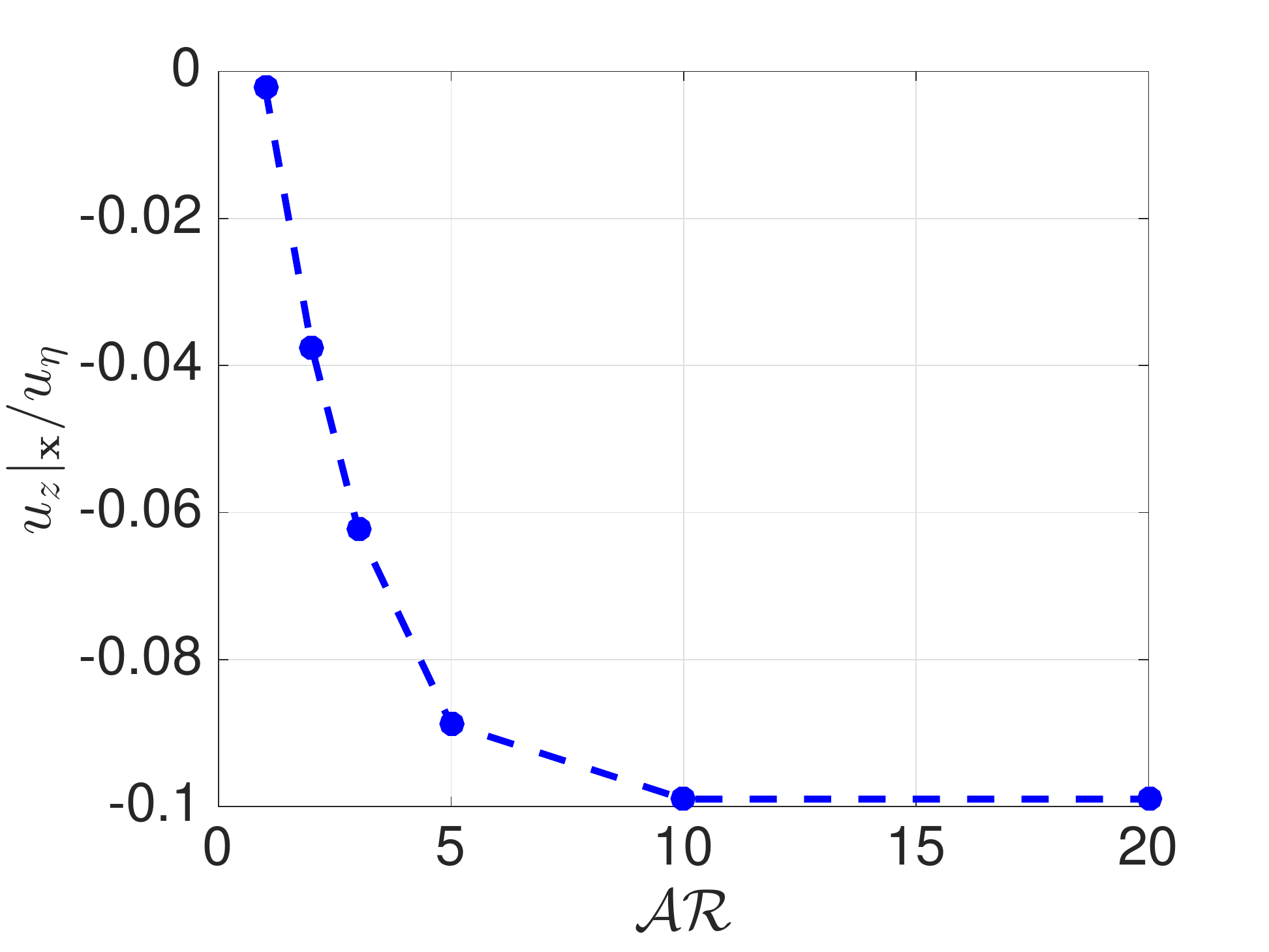}
\put(-395,120){$(a)$}
\put(-192,120){$(b)$} \\
\includegraphics[width=0.495\linewidth]{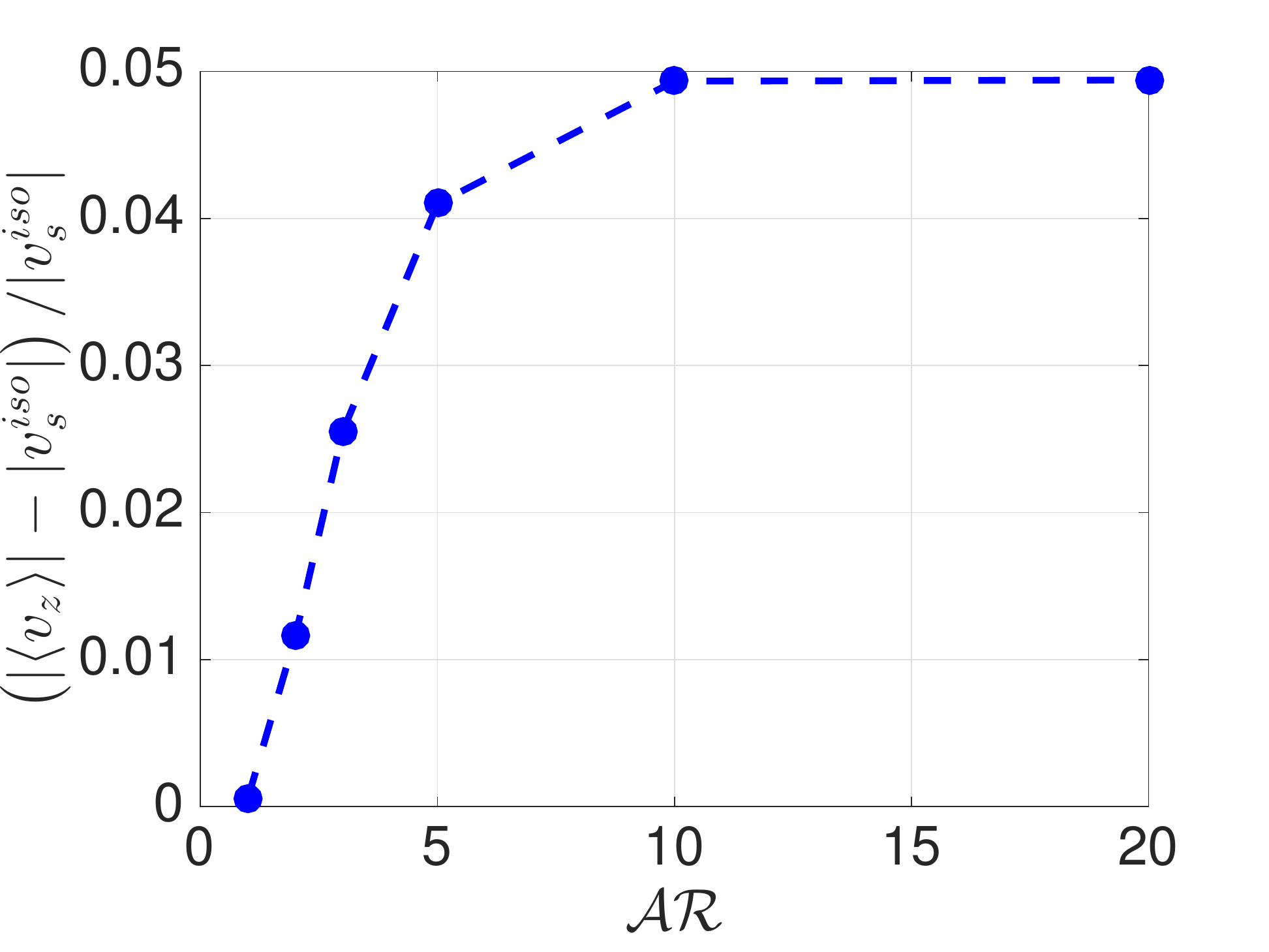}
\put(-204,120){\footnotesize (c)}
\caption{\label{fig:fix_vol} 
$(a)$ Radial distribution function of particle positions; the inset in this panel reports the fractal dimension of the clustering, estimated by fitting r.d.f. with a power-law in the range $r/\eta = [0.2:2]$, $(b)$ average of fluid vertical velocity at the particle position and $(c)$ increase in the settling velocity (difference in the settling rate with respect to the quiescent settling average $v^{iso}_s$, normalized by $v^{iso}_s$).} 
\end{figure}

We first investigate the settling of prolate spheroidal particles of constant volume, focusing on the effect of the particle aspect ratio.
The parameters used are $Re_\lambda = 100$, $\frac{\rho_p}{\rho_f} = 1.05$ and $D_{eq} = \eta/6$ where $D_{eq}$ is the diameter of a sphere having the same volume as the prolate spheroid.  Fixing the volume, the settling speed varies with the aspect ratio. It should be noted that for the higher aspect ratio cases ($\mathcal{AR}\geq15$), examined here, the larger diameter of the spheroid exceeds Kolmogorov length scale $\eta$ ($\approx 1.24$ for $\mathcal{AR} = 24$). This barely affects the results since rods act as tracers when their length is less than $5\eta$ and deviations from tracer behavior are very small until about $15\eta$ \citep{Parsa2014}. For the simulations presented here, the average settling speed of an uniformly distributed suspension, $v^{iso}_s$, varies in the range $1.5-3 u_\eta$. Table~\ref{table:Sediment} shows the settling velocities for the investigated cases. The reported values are all normalized by the Kolmogorov velocity. Note that the average settling speed $v^{iso}_s$ for spheroids in a quiescent flow is calculated as:    

\begin{equation}
v^{iso}_s = \frac{\left( v^{max}_s + 2 v^{min}_s \right) }{3} \, ,
\label{eq:avergaevelocity}  
\end{equation}
based on the assumption that the orientation of spheroids (cosine of the angle between spheroid's symmetric axis and the gravity direction) are uniformly distributed. 

The results are shown in figure~\ref{fig:fix_vol}. The main observation is that within the parameter space relevant to diatom chains, settling spheroids weakly cluster (figure~\ref{fig:fix_vol} panel a) and preferentially sample regions of down-welling flow, leading to downward velocity biases of up to $10\%$ of the Kolmogorov velocity $u_\eta$ (figure~\ref{fig:fix_vol} panel b). This corresponds to an increase of the mean settling speed of about $5\%$ compared to the mean settling speed in quiescent fluid (panel c in figure~\ref{fig:fix_vol}). These effects increase with the aspect ratio, $\mathcal{AR}$, and saturate for $\mathcal{AR}\ge10$.

\begin{figure}
\centering
\includegraphics[width=0.495\textwidth]{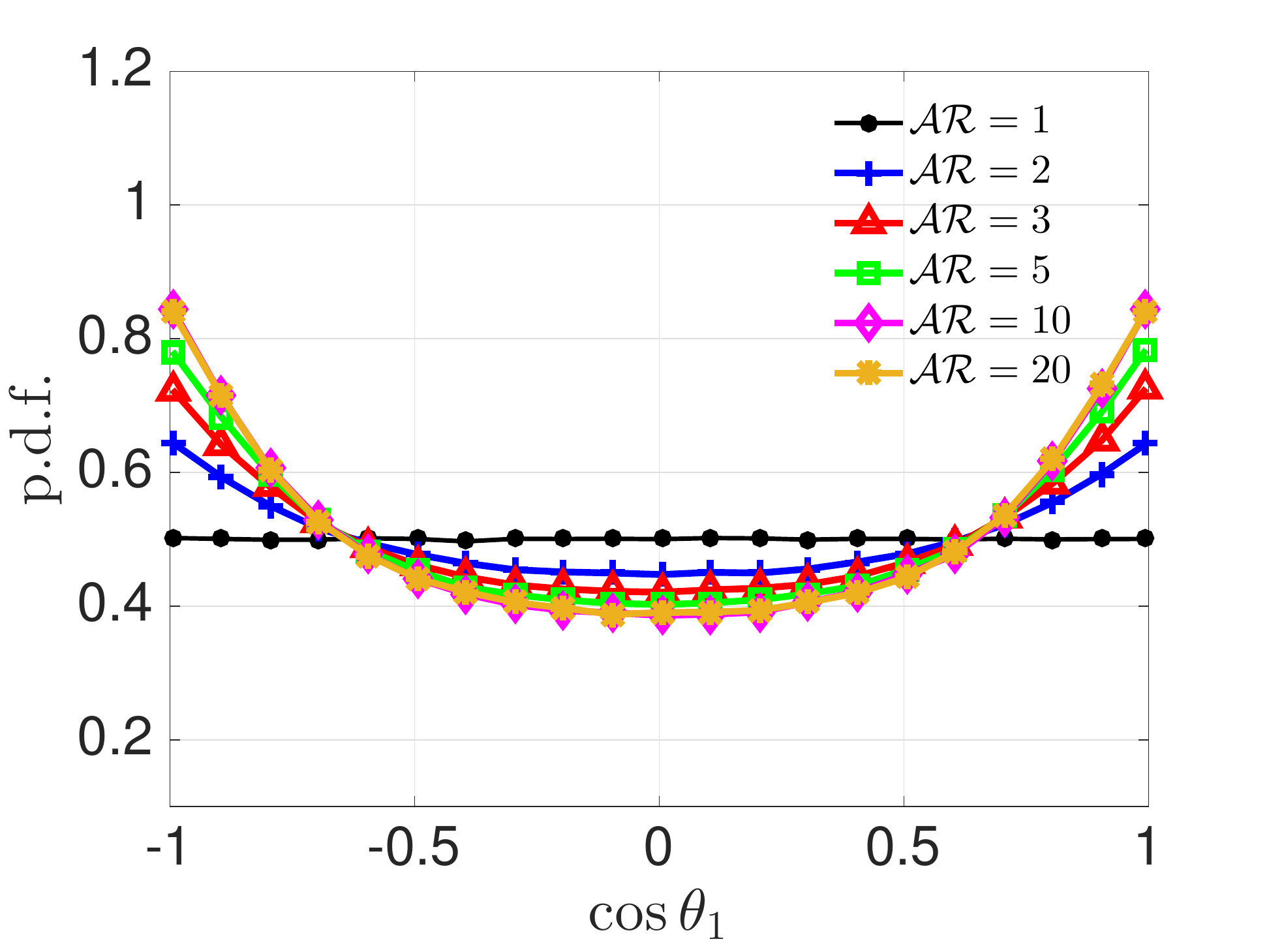}
\includegraphics[width=0.495\textwidth]{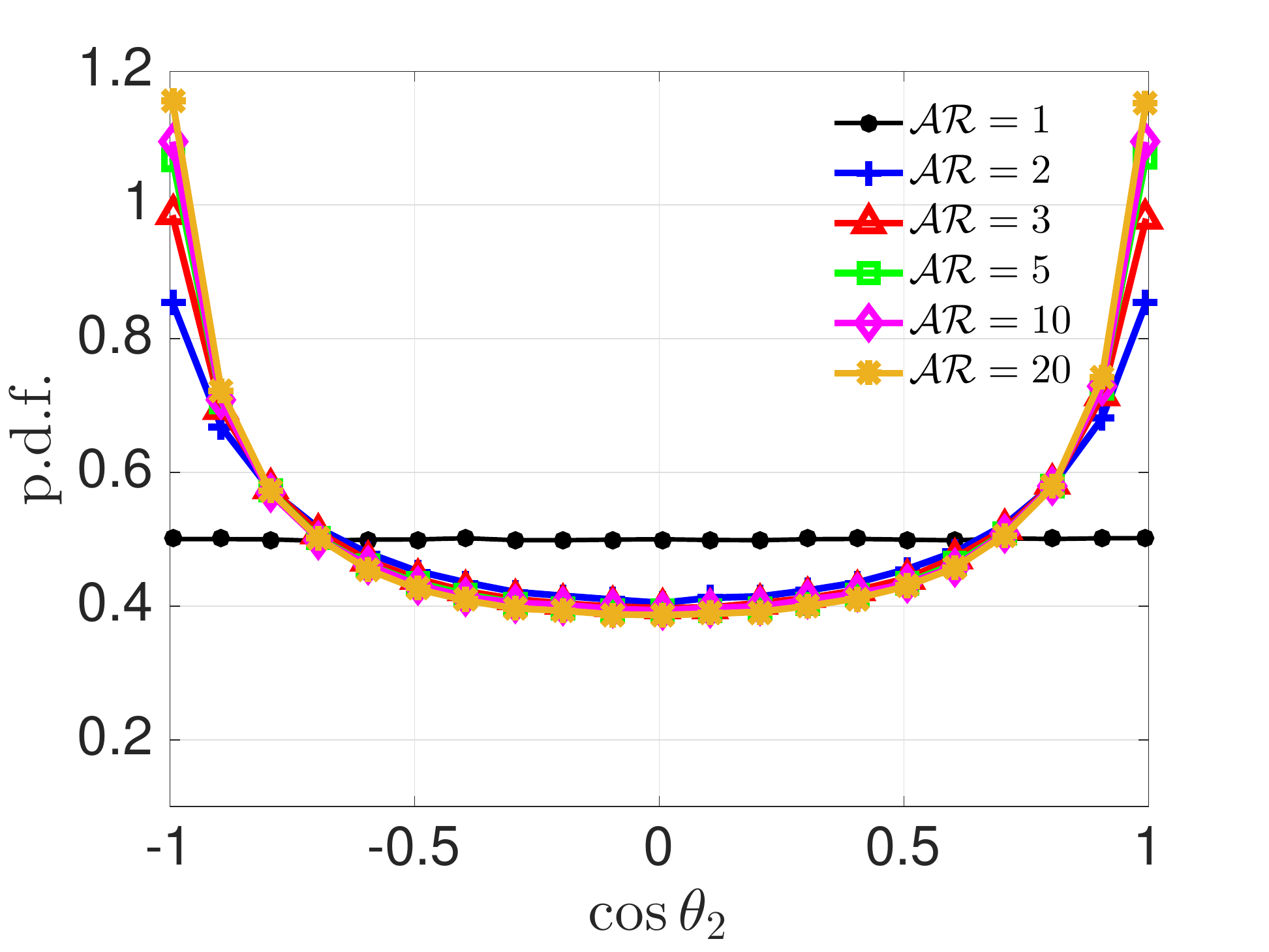}  
\put(-395,120){$(a)$}
\put(-197,120){$(b)$} \\
\includegraphics[width=0.495\textwidth]{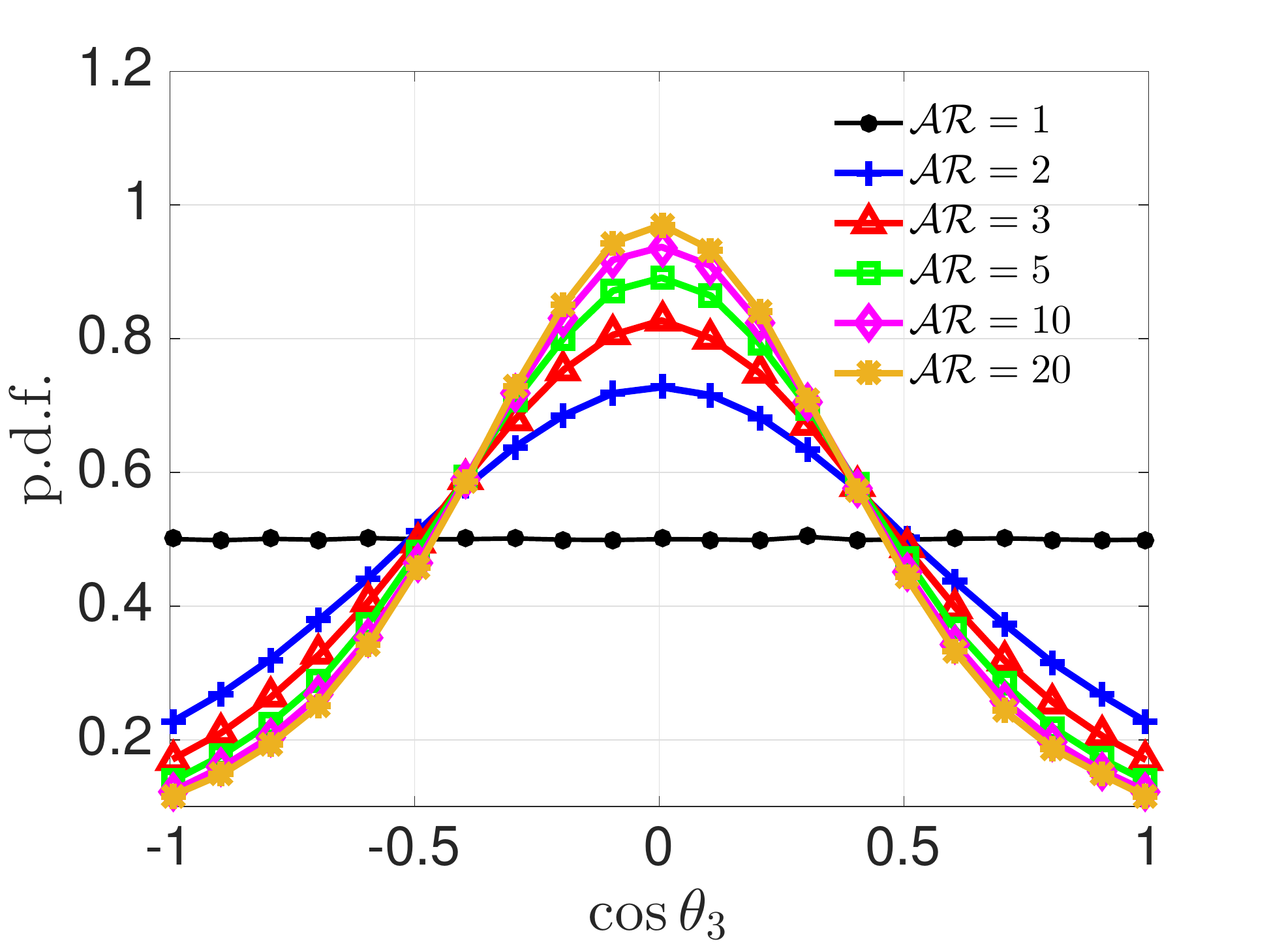}
\includegraphics[width=0.495\textwidth]{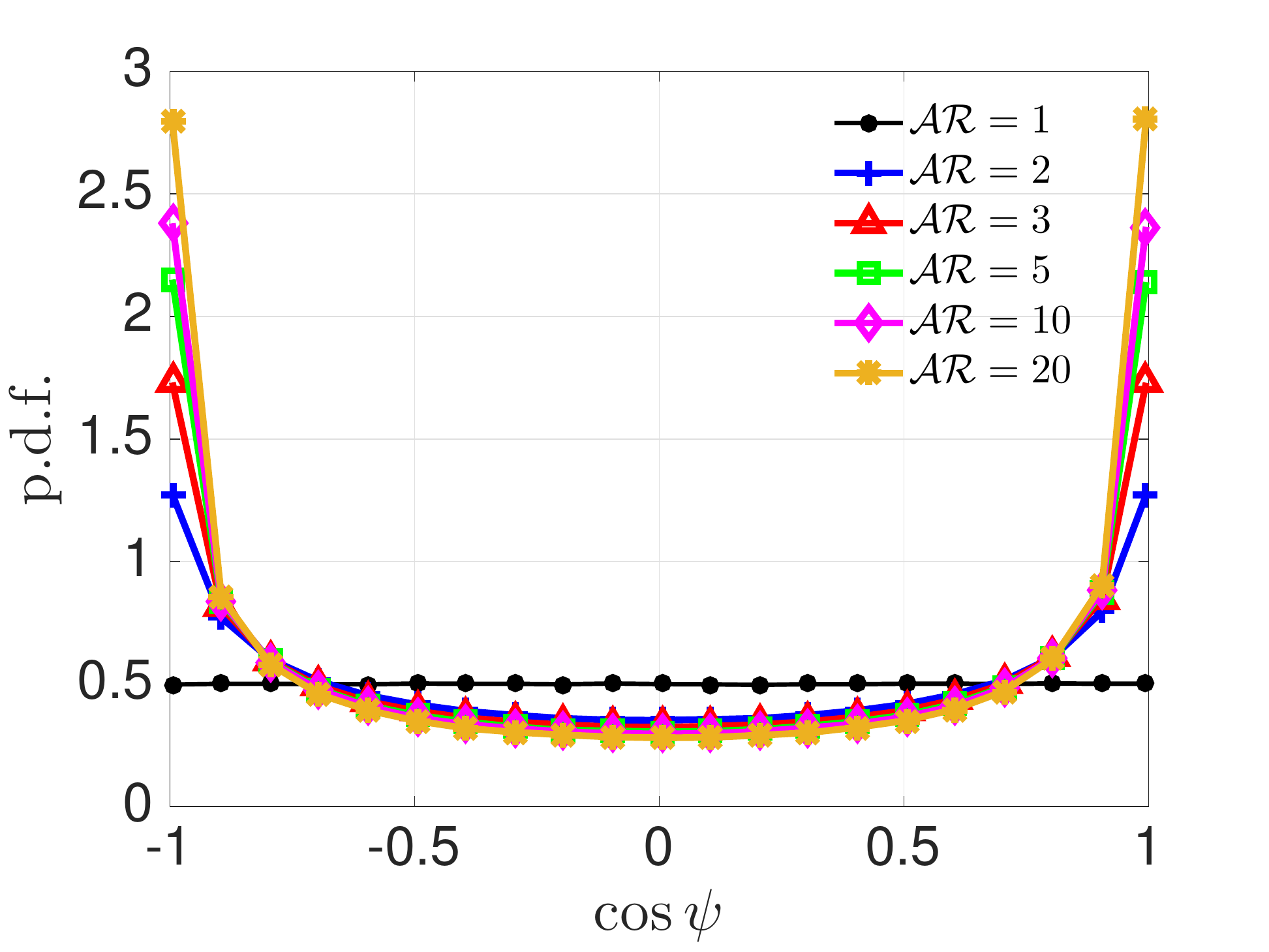}
\put(-395,120){$(c)$}
\put(-197,120){$(d)$} \\
\caption{\label{fig:orient}  
p.d.f of orientation of particles with different aspect ratios with respect to the three eigenvectors of the deformation tensor $(a-c)$ and $(d)$ the local vorticity vector.} 
\end{figure}

To quantify patchiness (clustering) in fully three-dimensional isotropic turbulent flows,  we use the radial pair distribution function (RDF), sometimes also called the correlation function.  The RDF measures the probability of finding a particle pair separated by a given radial distance, normalized by the value expected from a uniform distribution.
This is defined as
\begin{equation}
g\left(r\right)=\frac{1}{4\pi r^2} \frac{dN_{r}}{dr} \frac{1}{n_{0}},
\label{eq5}
\end{equation}
where, $n_{0}=0.5N_{p}\left(N_{p}-1\right)/V_{0}$ is the density of pairs in the whole volume $V_{0}$. $N_{p}$ is the total number of particles in the domain and $N_{r}$ is the number of pairs at distance $r$. The small volume fraction, used here, removes the exclusion effects, and thus an RDF for a random distribution results in a flat value of $1$. The radial distribution function (RDF) is reported  in figure~\ref{fig:fix_vol}(a). 
In the inset of the same figure, we report  
the scaling exponent of RDF at small separations, also often used as an indicator of patchiness \citep{Durham2013}. 
This exponent denotes the fractal dimension of the set where particles are found. Patchiness is seen to increase with the particle aspect ratio and saturate at the largest values considered.

In figure~\ref{fig:orient}, we study  the relation between particle orientation and the underlying flow field when varying the particle shape. 
In figure~\ref{fig:orient}(a), (b) and (c), we show the orientation of the particle with respect to the three eigendirections of the strain tensor,  defined such that $\lambda_1> \lambda_2> \lambda_3$ so that the first eigendirection is extensional and the third eigendirection is compressional. The angles between the eigendirections and the spheroid orientation are denoted as $\theta_1$, $\theta_2$ and $\theta_3$.
 In figure~\ref{fig:orient}d, we show $\psi$, the angle between the orientation and the local vorticity vector. 
 While spherical particles, $\mathcal{AR}=1$, are isotropic in nature, elongated spheroids tend to align with the local flow strain, more frequently with the second eigendirection, and with the local flow vorticity.
 Indeed, the pdfs peak when prolate particles are parallel to the eigendirections associated with $\lambda_2$ and $\lambda_1$. An increasing probability of parallel alignment with the first two eigendirections of the strain is seen with increased aspect ratio.  The spheroid is most likely to be normal to the third eigendirection of the strain as shown in figure~\ref{fig:orient}(c). 
The strongest tendency to align with the local vorticity vector is consistent with previous observations for neutrally buoyant spheroids \citep{Chevillard2013,Ni2015}. 
The alignment with the vorticity vector increases with the aspect ratio $\mathcal{AR}$.

\subsection{Numerical experiments}
 
To understand how the observed increase of settling speed and preferential sampling depends on the different governing parameters, we perform a series of numerical experiments in order to investigate the effect of the aspect ratio, $\mathcal{AR}$, the amplitude of the sedimenting velocity $v^{iso}_s$ and the ratio between the maximum (spheroid aligned with the gravity direction) and minimum (normal to gravity) settling speed $v_s^{max} / v_s^{min}$. In each series of simulations we fix all the parameters except for one to focus on a single parameter effect. While varying only one parameter can result in unrealistic conditions occasionally, it sheds light on the isolated effect of the parameter of interest on particles' collective behavior.  As mentioned above, the non-inertial (kinematic) model is used in these simulations.

\subsubsection{Effect of sedimenting velocity magnitude}

\begin{figure}
\centering
\includegraphics[width=0.495\linewidth]{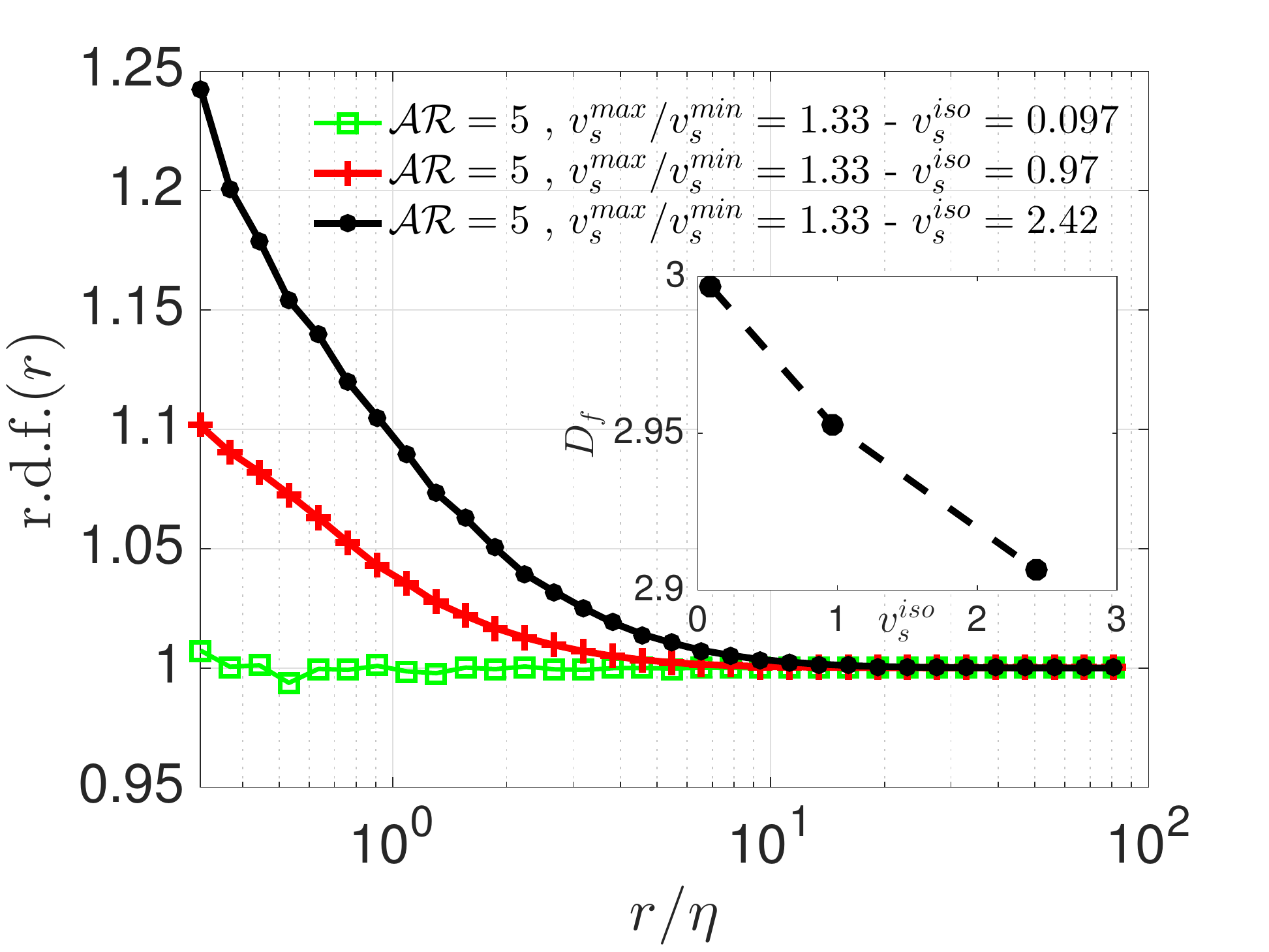}
\includegraphics[width=0.495\linewidth]{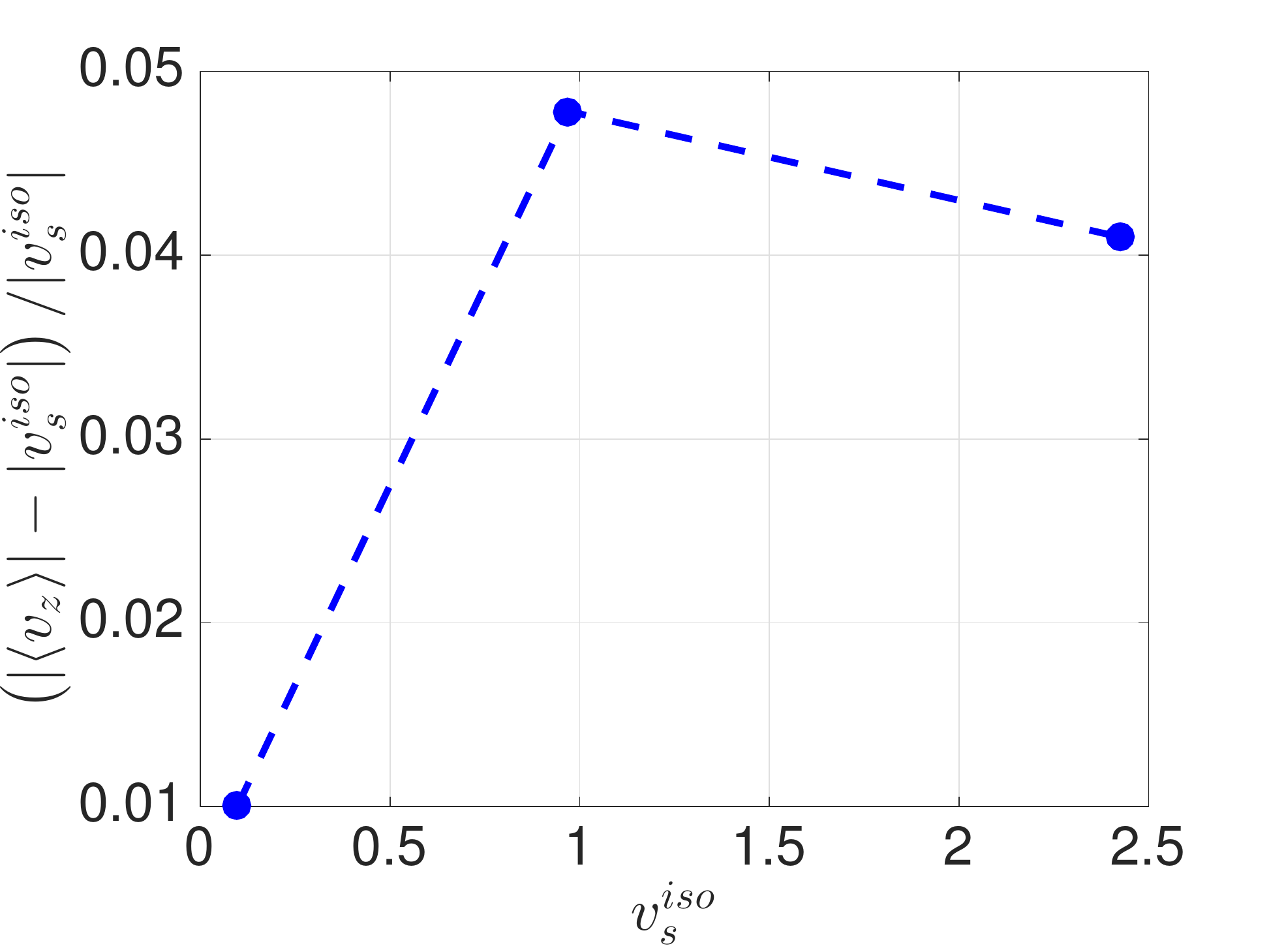}
\put(-395,123){$(a)$}
\put(-200,123){$(b)$}
\caption{\label{fig:vel-mag} 
$(a)$ Radial distribution function of particle positions and $(b)$ increase in the settling velocity for $\mathcal{AR}=5$ and the indicated values of $v_s^{iso}$.} 
\end{figure}

First we examine how particle clustering varies with the amplitude of the settling speed. To this end, we keep the aspect ratio $\mathcal{AR}=5$, corresponding to a ratio between maximum and minimum settling speed, $v_s^{max}/v_s^{min} = 1.33$, and change $v_s^{iso}$ from 0.1 to about 2.5 Kolmogorov velocities (this may correspond to a change in the particle size with respect to the turbulence Kolmogorov scale or a change in particle density). Results in figure~\ref{fig:vel-mag}(a) show that patchiness increases with the average settling speed, as measured by the RDF or by the fractal dimension (inset in figure~\ref{fig:vel-mag}(a)). The increased clustering, however, does not correspond to an increase of the relative average settling speed when $v_s^{iso}$ is greater than the Kolmogorov velocity.

\subsubsection{Effect of aspect ratio}

\begin{figure}
\centering
\includegraphics[width=0.495\linewidth]{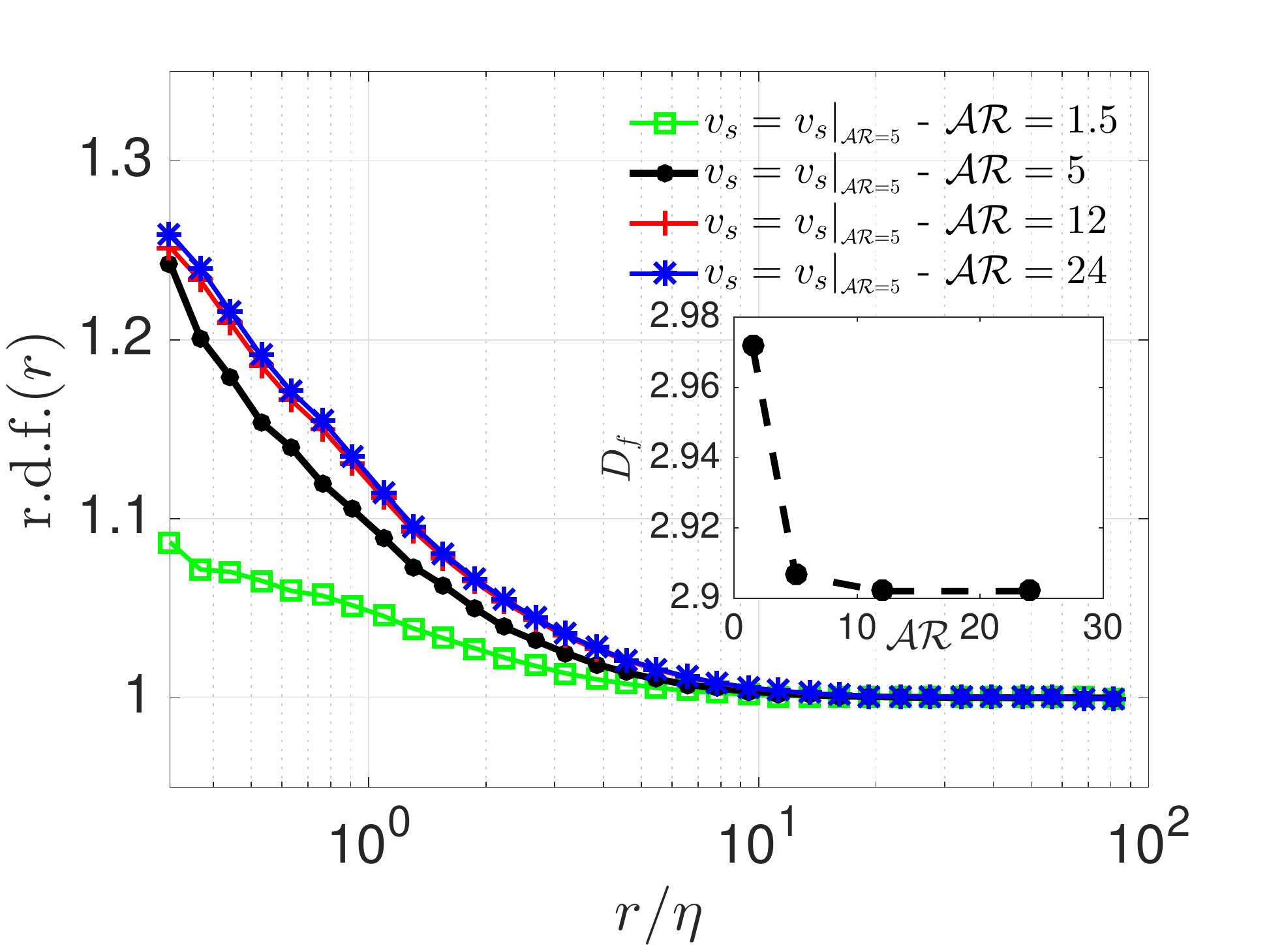}
\includegraphics[width=0.495\linewidth]{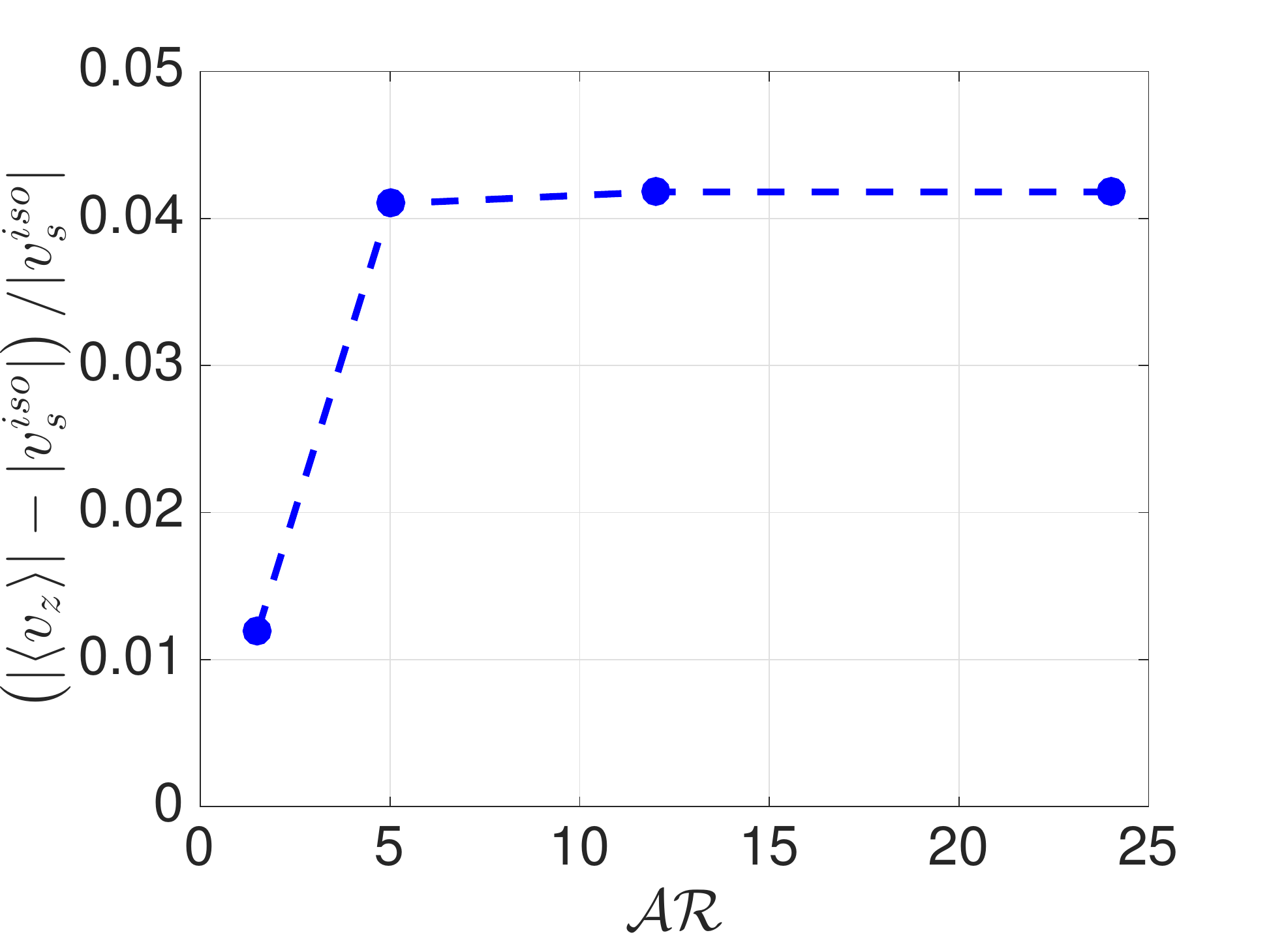}
\put(-395,123){$(a)$}
\put(-200,123){$(b)$}
\caption{\label{fig:diffAR} 
$(a)$ Radial distribution function of particle positions and $(b)$ increase in the settling velocity for the different aspect ratios indicated, while keeping the same maximum and minimum settling speed.} 
\end{figure}

To elucidate the role of the particle rotation rate on the observed preferential sampling, we vary $\mathcal{AR}$, while artificially keeping the settling speed fixed, both the components parallel and normal to gravity, fixed at values corresponding to $AR=5$ in table~\ref{table:Sediment}. The solution of Jeffery's equation (\ref{eq:RotationTracers}) reveals that the time spent aligned with the local shear increases with increasing aspect ratio. The results, displayed in figure~\ref{fig:diffAR}, show that the RDF and the relative increase of sedimenting velocity saturate for $\mathcal{AR} \geq 5$. Interestingly, rotation rates of oblate and prolate particles are seen to saturate for $\mathcal{AR}>5$  and  for $\mathcal{AR}<1/5$ in the study of \cite{Byron2015}. 
We have also observed, not shown here, that the alignment with the local vorticity increases with the aspect ratio and saturates when $\mathcal{AR}\geq5$ unlike the cases in figure~\ref{fig:orient}, where it continues to increase; this suggests that the alignment with vorticity is influenced by the difference between settling parallel and normal to gravity, which is fixed here.

\subsubsection{Effect of local alignment}

\begin{figure}
\centering
\includegraphics[width=0.6\textwidth]{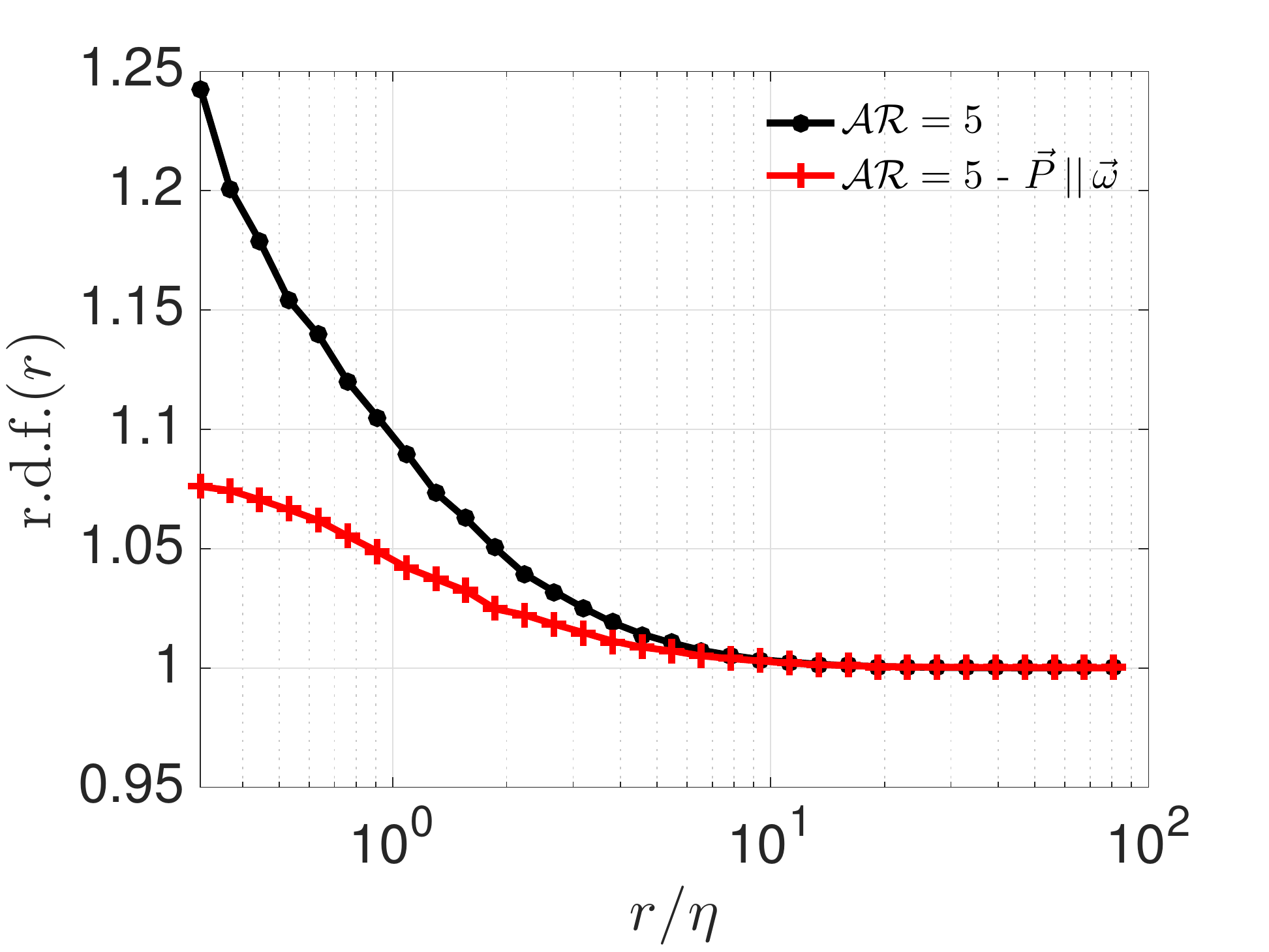}
\caption{\label{fig:vort}  
Radial distribution function of particle positions for particles with $\mathcal{AR}=5$ and particles with the same sedimenting velocity but with their orientation forced to be always aligned with the local vorticity vector.} 
\end{figure}

To show the isolated effect of particle alignment with the local vorticity vector we performed a simulation, forcing the elongated spheroid to be always aligned with the local vorticity (figure~\ref{fig:vort}): we observe that clustering decreases in this case, relative to the case of in which the particle was free to rotate, and therefore clustering cannot be explained by the alignment with the local flow vorticity.

\subsubsection{Effect of drag anisotropy} 

\begin{figure}
\centering
\includegraphics[width=0.495\linewidth]{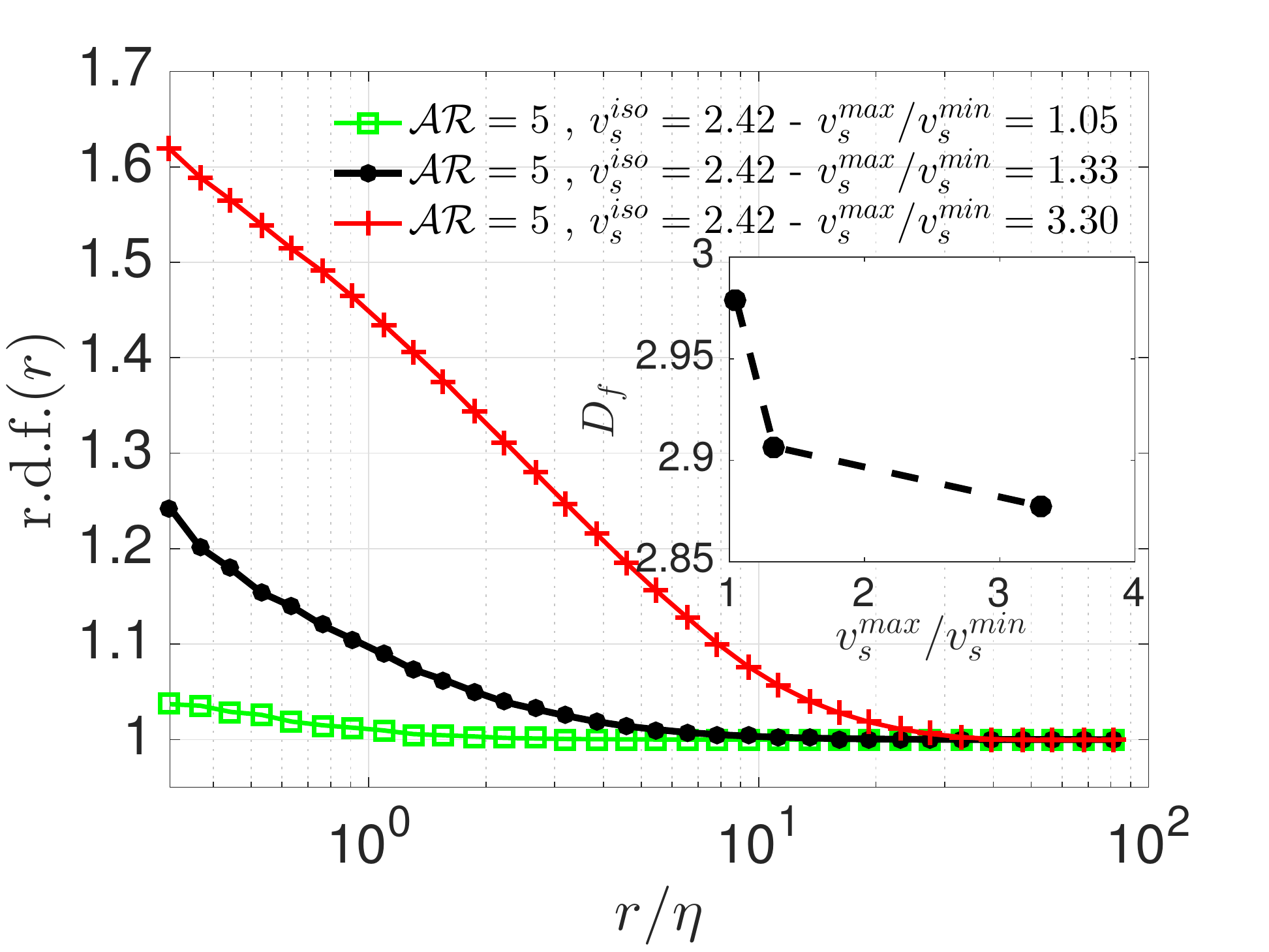}
\includegraphics[width=0.495\linewidth]{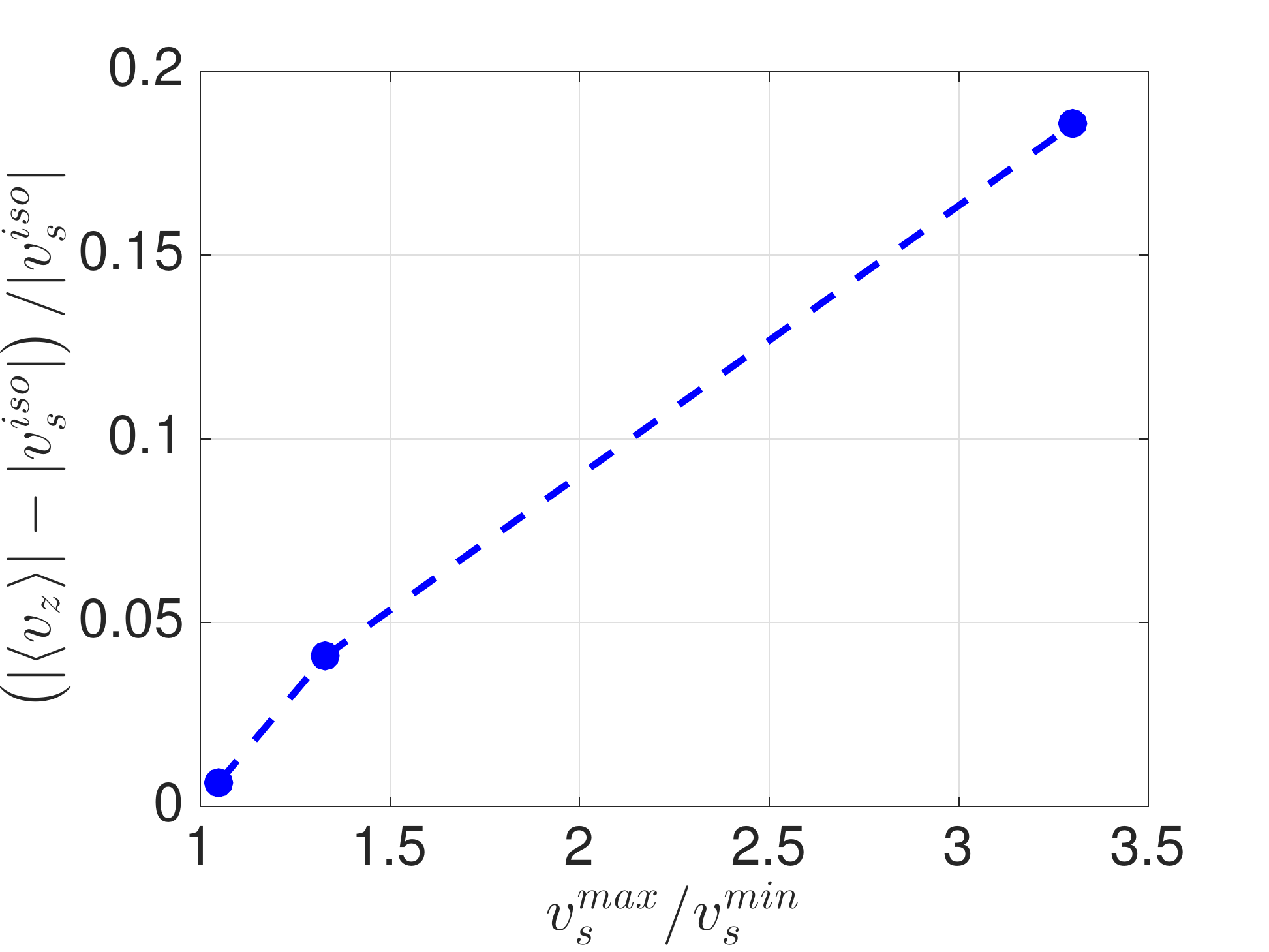}
\put(-395,123){$(a)$}
\put(-200,123){$(b)$}
\caption{\label{fig:aniso} 
$(a)$ Radial distribution function of particle positions and $(b)$ increase in the settling velocity for settling spheroids with aspect ratio $\mathcal{AR}=5$, fixed average settling speed $v_s^{iso} = 2.42$ and different ratios between settling speed in the direction normal and parallel to gravity, $v_s^{max} / v_s^{min}$, as indicated in the legend.} 
\end{figure}

\begin{figure}
\centering
\includegraphics[width=0.6\textwidth]{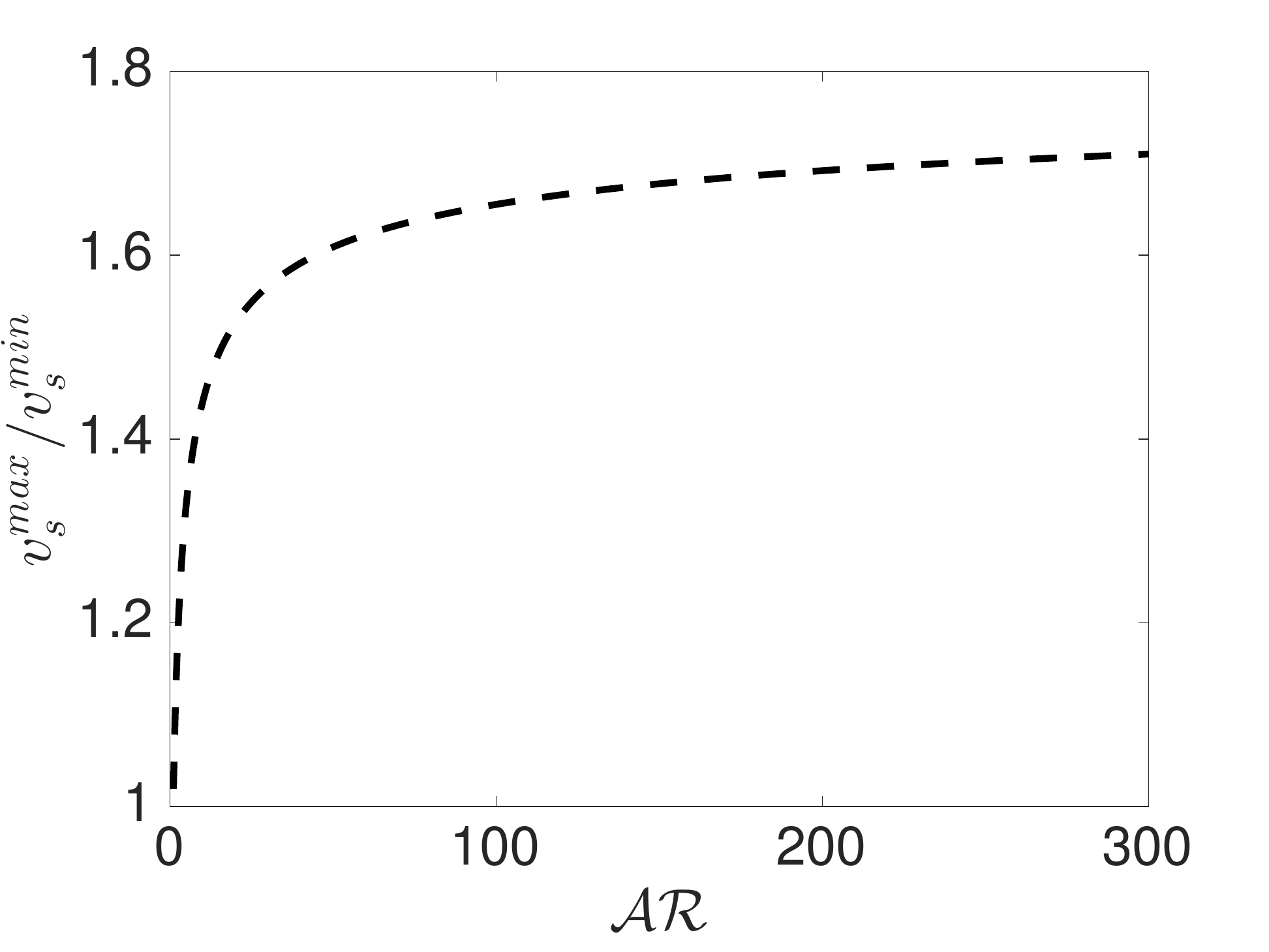}
\caption{\label{fig:ratioSaturation}  
$v_s^{max} / v_s^{min}$ versus $\mathcal{AR}$ for prolate particles, see e.g.\  analytical expressions reported in \cite{Dahlkild2011}.} 
\end{figure}

One important feature of non-spherical particles in inertialess flows is their drag anisotropy, which can also be exploited by micro-organisms for locomotion. In the case of settling prolate spheroids, this anisotropy is seen in the oblique settling of inclined particles and in the difference between settling speed when oriented parallel or normal to gravity. 
For the results presented so far we followed the analytical expression for the settling speed derived for prolate spheroids. Shapes in nature are, however, more complicated. Flexibility and non-uniformly distributed mass are likely to cause deviation from the idealized case. To understand more the relevance of drag anisotropy we perform simulations of spheroids with aspect ratio $\mathcal{AR}=5$ and fixed average settling speed $v_s^{iso} = 2.42$, while varying the ratio between settling speed in the direction normal and parallel to gravity, $v_s^{max} / v_s^{min}$.
As seen in figure~\ref{fig:aniso}(a) clustering increases significantly with drag anisotropy, or $v_s^{max} / v_s^{min}$. More interestingly, the relative settling speed increases by up to about $20\%$ when $v_s^{max} / v_s^{min}=3.3$. The value $3.3$ is chosen to artificially emphasize the drag anisotropy effect; however, note that the ratio between maximum and minimum settling speed of spheroidal particles increases with the aspect ratio and saturates at about $1.75$ (figure~\ref{fig:ratioSaturation}).

\subsection{Mechanism for increased settling speed}

\begin{figure} 
\includegraphics[width=0.98\textwidth]{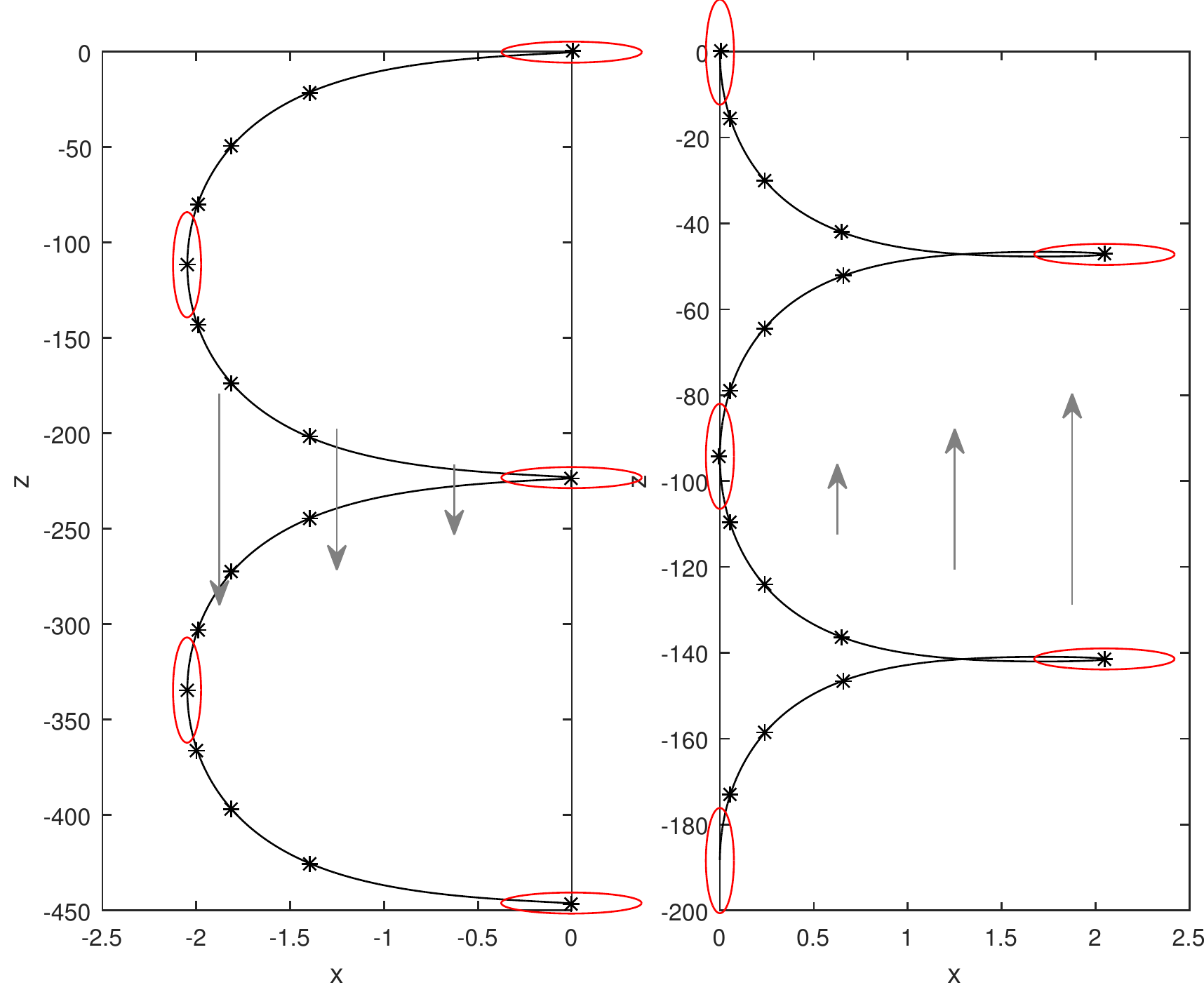}
\put(-375,280){$(a)$}
\put(-177,280){$(b)$}
\caption{\label{fig:JeffSed} 
Trajectory of single particle in vertical laminar shear over a Jeffery orbit. Particle initially at the origin with orientation (a) $\theta=\pi/2$, (b) $\theta=0$. Stars indicate position at equally spaced time intervals, red ellipses indicate orientation of particle orientation. Grey arrows indicate flow velocity  $\mathbf{u}=-\gamma x  \hat{\mathbf{e}}_g$. }
\end{figure}

To understand why ellipsoidal particles preferentially locate in regions of down-welling flow and thus have an enhanced settling speed, we consider vertical linear shear, $\mathbf{u}=-\gamma x  \hat{\mathbf{e}}_g$ with motion constrained to the $x-z$ plane. We define $\theta$ to be the angle the particle symmetry axis makes with the vertical, so that in $x-z$ co-ordinates the orientation vector is given by  $\mathbf{p}=(\sin\theta,\cos\theta)^T$. 

From Jeffery's equation (\ref{eq:RotationTracers}) the time evolution of particle orientation is given by  
\begin{eqnarray}
\dot{\theta}=\frac{\gamma}{2}\left(-1+\alpha \cos 2\theta\right) .
\end{eqnarray}
Note that the particle rotates most slowly when $\theta=0$ or $\pi$, and thus spends more time in these states, i.e. aligned with the local shear.
From equation  (\ref{eq:VelocityTracers}), the vertical velocity of a particle is given by
\begin{eqnarray*}
v_z=\gamma x -v_s^{min}- \left (v_s^{max}-v_s^{min}\right ) \cos^2\theta.
\end{eqnarray*}
The time-averaged vertical velocity over a Jeffery orbit (of duration $T=4\pi/\gamma \sqrt{1-\alpha^2}$) is given by  
 \begin{eqnarray*}
\overline{v_z}&=&\frac{1}{T}
\int_0^T(\gamma x -v_s^{min}- \left (v_s^{max}-v_s^{min}\right ) \cos^2\theta)dt\\
\nonumber
&=& \gamma \overline{x} -v_s^{min}- \frac{ \left (v_s^{max}-v_s^{min}\right ) \sqrt{1-\alpha^2}}{2\pi }\int_0^{2\pi}\frac{\cos^2\theta}{-1+\alpha \cos 2\theta}   d\theta,
\end{eqnarray*}
where $\overline{x}$ is the time-averaged horizontal position. The effect of the shear strength, $\gamma$, only appears through the term $\gamma \overline{x}$, and so the effect of the shear on the total vertical velocity of particles can be determined if we know the time-averaged horizontal position. In particular, the only way cells are predicted to exhibit enhanced settling is if they spend more time in down-welling regions.

To demonstrate how the time-averaged horizontal position is biased towards down-welling regions, yielding enhanced settling,  we numerically compute particle trajectories for a complete Jeffery orbit for an example set of parameters taken from table~\ref{table:Sediment}: $\gamma=1$, $\mathcal{AR}=20$ ($\alpha = 0.995$),  $v_s^{min}=1.29$ and $v_s^{max}=1.97$ (see figure~\ref{fig:JeffSed}). Particles are initially located at the origin, and the initial orientation is taken as $\theta=0$ or $\theta=\pi/2$. Particles initially horizontal ($\theta=\pi/2$, figure~\ref{fig:JeffSed}a) have a negative horizontal component due to shear-induced lift, causing them to translate into down-welling flow. As the particle rotates, it reaches a minimum horizontal position at the point when $\theta=0$ and then reverses its horizontal component of motion back towards $x=0$ which is attained when $\theta=3\pi/2$. The particle has a slowest rotation rate when $\theta=0$ or $\pi$  corresponding to the minimum horizontal position, and thus the time-averaged horizontal position is less than the midpoint of the horizontal limits of the trajectory. This bias represents a tendency to cluster in down-welling flow. Particles initially vertical ($\theta=0$, figure~\ref{fig:JeffSed}b) sediment with a positive horizontal component into upwelling flow. As the particle rotates, it reaches a maximum horizontal position at the point when $\theta=3\pi/2$ and then reverses its horizontal component of motion back towards $x=0$ which is attained when $\theta=\pi$. Again, the particle has a slowest rotation rate at the minimum horizontal position, and thus the time-averaged horizontal position is less than the midpoint of the horizontal limits of the trajectory. This simple example shows how shear, rotation, lift, and drag-anistropy couple to bias particle trajectories towards downward flow.

\section{Final remarks} 
\label{sec:Final_remarks} 

We report results from simulations of elongated, rigid particles (prolate spheroids) with different aspect ratios sinking in homogenous, isotropic turbulence. Parameters used in this study represent marine or freshwater diatoms that are characterized by very small relaxation times compared to that of the fluid but with high settling velocities on the order of Kolmogorov velocities.  Given the small size of diatoms relative to the Kolmogorov length-scale and their short relaxation time, cells are assumed to behave like passive tracers with a correction given by their Stokes' settling velocity. This assumption is confirmed by comparing with a dynamic model where Stokes' drag, gradient pressure and added mass are taken into account. Results from these models show the same dynamics in the range of Stokes numbers and density ratios relevant to diatoms. We therefore use the kinematic (non-inertial tracer) model, which allows simpler control over different parameters, allowing us to study how and why turbulence affects sinking velocities and clustering of diatoms.

\cite{Ruiz2004} showed that turbulence increases the average sinking velocity of the diatom \textit{Coscinodiscus sp.} in a laboratory set-up. They attributed their observations to preferential sampling of down-welling regions of the flow, a mechanism that was previously investigated for highly inertial particles with Stokes numbers that are several orders of magnitude larger than those of phytoplankton, and hypothesized that preferential sampling will also result in local clustering.  Here we show numerically that preferential sampling occurs within the parameter space relevant to diatoms but only for non-spherical particles.  Preferential sampling of down-welling flow results in local clustering of particles and a relative increase in settling velocity, but the effect on the latter is smaller ($\le 5\%$), compared to the one shown in the lab \citep{Ruiz2004}. We provide further insight on the mechanism involved by showing that elongated spheroids tend to align with the local flow strain (more frequently with the second eigendirection) and with the local flow vorticity  \citep[this is in agreement with previous studies on neutrally buoyant spheroids in homogenous isotropic turbulence,][]{Ni2015}. To understand why ellipsoidal particles preferentially locate in regions of down-welling flow and thus have an enhanced settling speed, we consider a simple case of one particle sedimenting in linear shear flow.  We demonstrate how the time-averaged (over a complete Jeffery orbit) horizontal position comes to be biased towards down-welling regions due to the interaction of orientation, shear, drag, and lift.

Cell shape is an important morphological trait affecting many aspects in the ecology of phytoplankton \citep{Smayda1970,Lewis1976,Sournia1982,Reynolds1989,Reynolds1997,Karp2016} and results from our numerical model further highlight the complex interactions of cell shape with ambient flows.  Interestingly, the observed increase in settling speeds does not increase indefinitely, and saturates for $\mathcal{AR} > 10$. Increasing the aspect ratio, while keeping the particle volume and density ratio constant, varies other characteristics of the particle simultaneously. For example, the mean settling speed in quiescent flow decreases as the ratio between maximum (when particle is oriented parallel to the gravity direction) and minimum (when particle is oriented normal to the gravity direction) settling speed increases.  We therefore perform a series of numerical experiments to isolate the effect of the aspect ratio $\mathcal{AR}$, the magnitude of settling velocity $v^{iso}_s$ and the ratio between the maximum and minimum settling speed $v_s^{max} / v_s^{min}$ in quiescent flow.  When $v^{iso}_s$ and $v_s^{max} / v_s^{min}$ are held constant, the increase in clustering and settling speed begins to saturates at $\mathcal{AR} = 5$. This indicates that the relative enhancement that we observe for ellipsoids with aspect ratios larger than $5$ is not due to a direct enhancement by the aspect ratio effect but due to an increase in the ratio between the maximum and minimum settling speed ($v_s^{max} / v_s^{min}$) that is associated with a change in particle shape ($\mathcal{AR}$, while volume remains constant). Interestingly, an analysis of typical aspect ratios among phytoplankton shows a peak in distribution at $\mathcal{AR} \sim  5$ \citep{Clavano2007,Karp2016}, but this may be a coincidence because in nature multiple processes (grazing, diffusion, mechanical breakage by shear) act simultaneously as selective pressures on shape and size in phytoplankton \citep{Karp2016}.  

Spheroids with higher average Stokes settling speeds ($v^{iso}_s$) show greater tendency to cluster (Figure~\ref{fig:vel-mag}). However, when $v^{iso}_s$ is greater than Kolmogorov velocity, the increase in clustering does not correspond to an increase in relative average settling speed.  Clustering increases significantly with drag anisotropy (i.e., the ratio $v_s^{max} / v_s^{min}$) in a way that does not appear to saturate. Drag anisotropy increases with increasing $\mathcal{AR}$ and forces the particles to sample regions of down-welling flow to a further extent, even when the isolated effect of aspect ratio is saturated. 
\section*{Acknowledgements}
\vspace{10pt}

MNA, GS and LB acknowledge support by the European Research Council Grant No. ERC-2013-CoG-616186, TRITOS and the Swedish Research Council grant Outstanding Young Researcher to LB. Computer time is provided by SNIC (Swedish National Infrastructure for Computing). We also thank the support from the COST Action MP1305: Flowing matter, with particular mention to the discussions held during the workshop "Microorganisms in Turbulent
Flows", Lorentz Centre, Leiden, the Netherlands. This work is also supported by the National Science Foundation (NSF) grant No. 1334365 to LKB and grants No. 1334788 and 1604026 to EAV.

\appendix
\section{Parameter space for phytoplankton} \label{app:Parameter_Space} 
\begin{table}
  \begin{center}
\def~{\hphantom{0}}
  \begin{tabular}{lccc}
   \multicolumn{1}{c}{~~Parameter~~~~~~} & \multicolumn{2}{c}{~~~~Values} &    \\
  \cline{1-4}  \\
      Kinetic energy dissipation rate ($\epsilon$)  & ~~~~~~~$10^{-10}$ - $10^{-4} \,m^2s^{-3}$ & \\ [1pt]
      Taylor Reynolds number ($Re_\lambda$) & ~~~~~$50 - 200$ & \\ [1pt]
      Sinking velocity of diatoms ($v_s^{iso}$)   & ~~~~~~~$2 \, \mu m \,s^{-1}$ - $ 1 \,mm \,s^{-1}$ & \\ [1pt]
      Velocity ratio ($v_s^{iso}/ u_{\eta}$) & ~~~~~$0.02$ - $11$ & \\ [1pt]
      Size ratio ($L / \eta$) & ~~~~~$0.003$ - $2$ & \\ [1pt]
      Density ratio ($\rho_p / \rho_f$)   & ~~~~$1$ - $1.3$ & \\ [1pt]
      Relaxation time of a cell ($\tau_p$)  & ~~~~~$ \sim 10^{-4} \, s$ & \\ [1pt]
      Relaxation time of fluid ($\tau_f$) & $\,\,\,\,\, \,\, \sim \,\, 1 \, s$ & \\ [1pt]
      Stokes number ($\tau_p/\tau_f$) & ~~~~~$ \sim 10^{-4}$ & 
  \end{tabular}
  \caption{A summary of the parameter values relevant to marine diatoms. The value of $\epsilon$ is considered $10^{-6}$ $m^2s^{-3}$ for calculating $\tau_f$ and $Re_\lambda$.}
  \label{tab:param}
  \end{center}
\end{table}

Dissipation rates of turbulent kinetic energy $\epsilon$ in the ocean vary greatly in space and time.  In the open ocean, typical values range from $10^{-10}$ to $10^{-7} \,m^2s^{-3}$, while coastal regions, especially those influenced by tidal fronts, experience higher dissipation rates, on the order of $10^{-7}$ to $10^{-4} \,m^2s^{-3}$ \citep{Smyth2000,Barton2014}. Within the water column, turbulence kinetic energy dissipation rate are higher at the surface and decrease with depth \citep{Smyth2000}.  Given the kinematic viscosity of seawater $\nu \sim 10^{-6} m^2s^{-1}$, characteristic Kolmogorov length scale $\eta$ in the ocean ranges from $ \sim 300 \,\mu  m$ to a few $mm$.  Generally speaking, diatoms are smaller than characteristic Kolmogorov length scales in the ocean, and range in size from a few $\mu m$ for small solitary cells to 100s of $\mu m$ in length for chain forming species \citep{Karp1998}. This size ratio may become order $1$ for very long chains in strong turbulence but this would represent an extreme case. Sinking velocities of diatoms have been measured in quiescent water by measuring the temporal change in concentration of diatoms at the top and bottom of a water column with a known height and calculating average (community) settling velocities.  Using this approach, settling velocities ($v_s^{iso}$) were found to range between $2 \, \mu m \,s^{-1}$ to  $ \sim 1 \,mm \,s^{-1}$ \cite[reviewed in][]{Smayda1970}.   

Characteristic velocities at dissipation scale in the ocean ($u_{\eta} = (\nu \epsilon)^{1/4}$) are in the order of $0.01$ - $0.1$ $mm\, s^{-1}$, resulting in velocity ratio $v_s^{iso}/ u_{\eta} \approx$ $1$ - $4$ (assuming maximum $v_s^{iso}$ of $1 \,mm \,s^{-1}$). Densities of marine diatoms $\rho_p$ have not been measured directly but estimated from Stokes equation for settling spheres, based on measured sinking velocities.  Densities were found to be in the range of $1.03$ - $1.15$ $g \,cm^{-3}$ for actively growing cells with higher values for cells in senescence \cite[$1.059$-$1.33$ $g \,cm^{-3}$;][]{Eppley1967}. Average density of surface seawater $\rho_f$ is $1.025$ $g\, cm^{-3}$ and thus the density ratio relevant to the parameter space of phytoplankton is in the range of $1$ - $1.3$. Table~\ref{tab:param} summarizes the parameter values relevant to marine diatoms.

\bibliographystyle{jfm}
\bibliography{biblio}

\begin{thebibliography}{48}
\expandafter\ifx\csname natexlab\endcsname\relax\def\natexlab#1{#1}\fi
\def\au#1{#1} \def\ed#1{#1} \def\yr#1{#1}\def\at#1{#1}\def\jt#1{\textit{#1}}
  \def\bt#1{#1}\def\bvol#1{\textbf{#1}} \def\vol#1{#1} \def\pg#1{#1}
  \def\publ#1{#1}\def\arxiv#1{#1}\def\org#1{#1}\def\st#1{\textit{#1}}

\bibitem[Barry {\em et~al.\/}(2015)Barry, Rusconi, Guasto \&
  Stocker]{Barry2015}
{\sc \au{Barry, M.~T.}, \au{Rusconi, R.}, \au{Guasto, J.~S.} \& \au{Stocker,
  R.}} \yr{2015}  \at{Shear-induced orientational dynamics and spatial
  heterogeneity in suspensions of motile phytoplankton}.  \jt{Journal of The
  Royal Society Interface}  \bvol{12}~(112),  \pg{20150791}.

\bibitem[Barton {\em et~al.\/}(2014)Barton, Ward, Williams \&
  Follows]{Barton2014}
{\sc \au{Barton, A.~D.}, \au{Ward, B.~A.}, \au{Williams, R.~G.} \& \au{Follows,
  M.~J.}} \yr{2014}  \at{The impact of fine-scale turbulence on phytoplankton
  community structure}.  \jt{Limnology and Oceanography: Fluids and
  Environments}  \bvol{4}~(1),  \pg{34--49}.

\bibitem[Bec {\em et~al.\/}(2014)Bec, Homann \& Ray]{Bec2014}
{\sc \au{Bec, J.}, \au{Homann, H.} \& \au{Ray, S.~S.}} \yr{2014}
  \at{Gravity-driven enhancement of heavy particle clustering in turbulent
  flow}.  \jt{Physical review letters}  \bvol{112}~(18),  \pg{184501}.

\bibitem[Brenner(1964)]{brenner}
{\sc \au{Brenner, H.}} \yr{1964}  \at{The stokes resistance of an arbitrary
  particleÑiv arbitrary fields of flow}.  \jt{Chemical Engineering Science}
  \bvol{19}~(10),  \pg{703--727}.

\bibitem[Byron {\em et~al.\/}(2015)Byron, Einarsson, Gustavsson, Voth, Mehlig
  \& Variano]{Byron2015}
{\sc \au{Byron, M.}, \au{Einarsson, J.}, \au{Gustavsson, K.}, \au{Voth, G.},
  \au{Mehlig, B.} \& \au{Variano, E.}} \yr{2015}  \at{Shape-dependence of
  particle rotation in isotropic turbulence}.  \jt{Physics of Fluids
  (1994-present)}  \bvol{27}~(3),  \pg{035101}.

\bibitem[Chevillard \& Meneveau(2013)]{Chevillard2013}
{\sc \au{Chevillard, L.} \& \au{Meneveau, C.}} \yr{2013}  \at{Orientation
  dynamics of small, triaxial--ellipsoidal particles in isotropic turbulence}.
  \jt{Journal of Fluid Mechanics}  \bvol{737},  \pg{571--596}.

\bibitem[Clavano {\em et~al.\/}(2007)Clavano, Boss \& Karp-Boss]{Clavano2007}
{\sc \au{Clavano, W.~R.}, \au{Boss, E.} \& \au{Karp-Boss, L.}} \yr{2007}
  \at{Inherent optical properties of non-spherical marine-like particles from
  theory to observation}.  \jt{Oceanogr Mar Biol, Annu Rev}  \bvol{45},
  \pg{1--38}.

\bibitem[Dahlkild(2011)]{Dahlkild2011}
{\sc \au{Dahlkild, A.~A.}} \yr{2011}  \at{Finite wavelength selection for the
  linear instability of a suspension of settling spheroids}.  \jt{Journal of
  Fluid Mechanics}  \bvol{689},  \pg{183--202}.

\bibitem[Daitche(2015)]{Daitche2015}
{\sc \au{Daitche, A.}} \yr{2015}  \at{On the role of the history force for
  inertial particles in turbulence}.  \jt{Journal of Fluid Mechanics}
  \bvol{782},  \pg{567--593}.

\bibitem[Denny(1993)]{Denny1993}
{\sc \au{Denny, M.~W.}} \yr{1993} {\em Air and water: the biology and physics
  of life's media\/}.  \publ{Princeton University Press}.

\bibitem[Durham {\em et~al.\/}(2013)Durham, Climent, Barry, De~Lillo, Boffetta,
  Cencini \& Stocker]{Durham2013}
{\sc \au{Durham, W.~M.}, \au{Climent, E.}, \au{Barry, M.}, \au{De~Lillo, F.},
  \au{Boffetta, G.}, \au{Cencini, M.} \& \au{Stocker, R.}} \yr{2013}
  \at{Turbulence drives microscale patches of motile phytoplankton}.
  \jt{Nature communications}  \bvol{4}.

\bibitem[Durham {\em et~al.\/}(2011)Durham, Climent \& Stocker]{Durham2011}
{\sc \au{Durham, W.~M.}, \au{Climent, E.} \& \au{Stocker, R.}} \yr{2011}
  \at{Gyrotaxis in a steady vortical flow}.  \jt{Physical Review Letters}
  \bvol{106}~(23),  \pg{238102}.

\bibitem[Eppley {\em et~al.\/}(1967)Eppley, Holmes \& Strickland]{Eppley1967}
{\sc \au{Eppley, R.~W.}, \au{Holmes, R.~W.} \& \au{Strickland, J. D.~H.}}
  \yr{1967}  \at{Sinking rates of marine phytoplankton measured with a
  fluorometer}.  \jt{Journal of Experimental Marine Biology and Ecology}
  \bvol{1}~(2),  \pg{191--208}.

\bibitem[Fornari {\em et~al.\/}(2016)Fornari, Picano \& Brandt]{Fornari2016}
{\sc \au{Fornari, W.}, \au{Picano, F.} \& \au{Brandt, L.}} \yr{2016}
  \at{Sedimentation of finite-size spheres in quiescent and turbulent
  environments}.  \jt{Journal of Fluid Mechanics}  \bvol{788},  \pg{640--669}.

\bibitem[Gallily \& Cohen(1979)]{Gallily1979}
{\sc \au{Gallily, I.} \& \au{Cohen, A.~H.}} \yr{1979}  \at{On the orderly
  nature of the motion of nonspherical aerosol particles. ii. inertial
  collision between a spherical large droplet and an axially symmetrical
  elongated particle}.  \jt{Journal of Colloid and Interface Science}
  \bvol{68}~(2),  \pg{338--356}.

\bibitem[Guazzelli \& Morris(2011)]{Guazzelli2011}
{\sc \au{Guazzelli, E.} \& \au{Morris, J.~F.}} \yr{2011} {\em A physical
  introduction to suspension dynamics\/}, ,  \vol{vol.~45}.  \publ{Cambridge
  University Press}.

\bibitem[Gustavsson {\em et~al.\/}(2016)Gustavsson, Berglund, Jonsson \&
  Mehlig]{Gustavsson2016}
{\sc \au{Gustavsson, K.}, \au{Berglund, F.}, \au{Jonsson, P.~R.} \& \au{Mehlig,
  B.}} \yr{2016}  \at{Preferential sampling and small-scale clustering of
  gyrotactic microswimmers in turbulence}.  \jt{Physical review letters}
  \bvol{116}~(10),  \pg{108104}.

\bibitem[Gustavsson {\em et~al.\/}(2014)Gustavsson, Vajedi \&
  Mehlig]{Gustavsson2014}
{\sc \au{Gustavsson, K.}, \au{Vajedi, S.} \& \au{Mehlig, B.}} \yr{2014}
  \at{Clustering of particles falling in a turbulent flow}.  \jt{Physical
  Review Letters}  \bvol{112}~(21),  \pg{214501}.

\bibitem[Jeffery(1922)]{jeff}
{\sc \au{Jeffery, G.~B.}} \yr{1922} The motion of ellipsoidal particles
  immersed in a viscous fluid.  \bt{In {\em Proceedings of the Royal Society of
  London A: Mathematical, Physical and Engineering Sciences\/}}, ,  \vol{vol.
  102},  \pg{pp. 161--179}. The Royal Society.

\bibitem[Karp-Boss \& Boss(2016)]{Karp2016}
{\sc \au{Karp-Boss, L.} \& \au{Boss, E.}} \yr{2016}  \at{The elongated, the
  squat and the spherical: Selective pressures for phytoplankton shape}.
  \bt{In {\em Aquatic Microbial Ecology and Biogeochemistry: A Dual
  Perspective\/}},  \pg{pp. 25--34}.  \publ{Springer}.

\bibitem[Karp-Boss \& Jumars(1998)]{Karp1998}
{\sc \au{Karp-Boss, L.} \& \au{Jumars, P.~A.}} \yr{1998}  \at{Motion of diatom
  chains in steady shear flow}.  \jt{OCEANOGRAPHY}  \bvol{43}~(8).

\bibitem[Ki{\o}rboe(2008)]{Kiorboe2008}
{\sc \au{Ki{\o}rboe, T.}} \yr{2008} {\em A mechanistic approach to plankton
  ecology\/}.  \publ{Princeton University Press}.

\bibitem[Lazier \& Mann(1989)]{Lazier1989}
{\sc \au{Lazier, J. R.~N.} \& \au{Mann, K.~H.}} \yr{1989}  \at{Turbulence and
  the diffusive layers around small organisms}.  \jt{Deep Sea Research Part A.
  Oceanographic Research Papers}  \bvol{36}~(11),  \pg{1721--1733}.

\bibitem[Lewis(1976)]{Lewis1976}
{\sc \au{Lewis, W.~M.}} \yr{1976}  \at{Surface/volume ratio: implications for
  phytoplankton morphology}.  \jt{Science}  \bvol{192}~(4242),  \pg{885--887}.

\bibitem[Lucci {\em et~al.\/}(2011)Lucci, Ferrante \& Elghobashi]{Lucci2011}
{\sc \au{Lucci, F.}, \au{Ferrante, A.} \& \au{Elghobashi, S.}} \yr{2011}
  \at{Is stokes number an appropriate indicator for turbulence modulation by
  particles of taylor-length-scale size?}  \jt{Physics of Fluids
  (1994-present)}  \bvol{23}~(2),  \pg{025101}.

\bibitem[Marchioli {\em et~al.\/}(2014)Marchioli, Zhao \& Andersson]{zhao}
{\sc \au{Marchioli, C.}, \au{Zhao, L.} \& \au{Andersson, H.~I.}} \yr{2014}
  \at{On the relative rotational motion between rigid fibers and fluid in
  turbulent channel flow}.  \jt{Physics of Fluids (1994-present)}
  \bvol{28}~(1),  \pg{013301}.

\bibitem[Maxey(1987)]{Maxey1987}
{\sc \au{Maxey, M.~R.}} \yr{1987}  \at{The gravitational settling of aerosol
  particles in homogeneous turbulence and random flow fields}.  \jt{Journal of
  Fluid Mechanics}  \bvol{174},  \pg{441--465}.

\bibitem[Murray(1970)]{Murray1970}
{\sc \au{Murray, S.~P.}} \yr{1970}  \at{Settling velocities and vertical
  diffusion of particles in turbulent water}.  \jt{Journal of geophysical
  research}  \bvol{75}~(9),  \pg{1647--1654}.

\bibitem[Nguyen {\em et~al.\/}(2011)Nguyen, Karp-Boss, Jumars \&
  Fauci]{Nguyen2011}
{\sc \au{Nguyen, H.}, \au{Karp-Boss, L.}, \au{Jumars, P.~A} \& \au{Fauci, L.}}
  \yr{2011}  \at{Hydrodynamic effects of spines: A different spin}.
  \jt{Limnology and Oceanography: Fluids and Environments}  \bvol{1}~(1),
  \pg{110--119}.

\bibitem[Ni {\em et~al.\/}(2015)Ni, Kramel, Ouellette \& Voth]{Ni2015}
{\sc \au{Ni, R.}, \au{Kramel, S.}, \au{Ouellette, N.~T} \& \au{Voth, G.~A.}}
  \yr{2015}  \at{Measurements of the coupling between the tumbling of rods and
  the velocity gradient tensor in turbulence}.  \jt{Journal of Fluid Mechanics}
   \bvol{766},  \pg{202--225}.

\bibitem[Nielsen(1993)]{Nielsen1993}
{\sc \au{Nielsen, P.}} \yr{1993}  \at{Turbulence effects on the settling of
  suspended particles}.  \jt{Journal of Sedimentary Research}  \bvol{63}~(5).

\bibitem[Nielsen(2007)]{Nielsen2007}
{\sc \au{Nielsen, P.}} \yr{2007}  \at{Mean and variance of the velocity of
  solid particles in turbulence}.  \bt{In {\em Particle-Laden Flow\/}},
  \pg{pp. 385--391}.  \publ{Springer}.

\bibitem[Olivieri(2013)]{Olivieri2013}
{\sc \au{Olivieri, S.}} \yr{2013} Analysis of the forces acting on particles in
  homogeneous isotropic turbulence.

\bibitem[Olivieri {\em et~al.\/}(2014)Olivieri, Picano, Sardina, Ludicone \&
  Brandt]{Olivieri2014}
{\sc \au{Olivieri, S.}, \au{Picano, F.}, \au{Sardina, G.}, \au{Ludicone, D.} \&
  \au{Brandt, L.}} \yr{2014}  \at{The effect of the basset history force on
  particle clustering in homogeneous and isotropic turbulence}.  \jt{Physics of
  Fluids (1994-present)}  \bvol{26}~(4),  \pg{041704}.

\bibitem[Parsa \& Voth(2014)]{Parsa2014}
{\sc \au{Parsa, S.} \& \au{Voth, G.~A.}} \yr{2014}  \at{Inertial range scaling
  in rotations of long rods in turbulence}.  \jt{Physical review letters}
  \bvol{112}~(2),  \pg{024501}.

\bibitem[Reynolds(1989)]{Reynolds1989}
{\sc \au{Reynolds, C.~S.}} \yr{1989}  \at{Physical determinants of
  phytoplankton succession}.  \bt{In {\em Plankton ecology\/}},  \pg{pp.
  9--56}.  \publ{Springer}.

\bibitem[Reynolds \& Irish(1997)]{Reynolds1997}
{\sc \au{Reynolds, C.~S.} \& \au{Irish, A.~E.}} \yr{1997}  \at{Modelling
  phytoplankton dynamics in lakes and reservoirs: the problem of in-situ growth
  rates}.  \jt{Hydrobiologia}  \bvol{349}~(1-3),  \pg{5--17}.

\bibitem[Rogallo(1981)]{rogallo}
{\sc \au{Rogallo, R.~S.}} \yr{1981}  \at{Numerical experiments in homogeneous
  turbulence}.  \jt{NASA STI/Recon Technical Report N}  \bvol{81},  \pg{31508}.

\bibitem[Ruiz {\em et~al.\/}(2004)Ruiz, Mac{\'\i}as \& Peters]{Ruiz2004}
{\sc \au{Ruiz, J.}, \au{Mac{\'\i}as, D.} \& \au{Peters, F.}} \yr{2004}
  \at{Turbulence increases the average settling velocity of phytoplankton
  cells}.  \jt{Proceedings of the National academy of Sciences of the United
  States of America}  \bvol{101}~(51),  \pg{17720--17724}.

\bibitem[Sardina {\em et~al.\/}(2015)Sardina, Picano, Brandt \&
  Caballero]{prlcloud}
{\sc \au{Sardina, G.}, \au{Picano, F.}, \au{Brandt, L.} \& \au{Caballero, R.}}
  \yr{2015}  \at{Continuous growth of droplet size variance due to condensation
  in turbulent clouds}.  \jt{Physical review letters}  \bvol{115}~(18),
  \pg{184501}.

\bibitem[Smayda(1970)]{Smayda1970}
{\sc \au{Smayda, T.~J.}} \yr{1970}  \at{The suspension and sinking of
  phytoplankton in the sea} .

\bibitem[Smyth \& Moum(2000)]{Smyth2000}
{\sc \au{Smyth, W.~D.} \& \au{Moum, J.~N.}} \yr{2000}  \bt{Ocean turbulence}.
  {\em Tech. Rep.\/}.  \org{Technical report, College of Oceanic and
  Atmospheric Sciences, Oregon State University}.

\bibitem[Sournia(1982)]{Sournia1982}
{\sc \au{Sournia, A.}} \yr{1982}  \at{Form and function in marine
  phytoplankton}.  \jt{Biological Reviews}  \bvol{57}~(3),  \pg{347--394}.

\bibitem[Tooby {\em et~al.\/}(1977)Tooby, Wick \& Isaacs]{Tooby1977}
{\sc \au{Tooby, P.~F.}, \au{Wick, G.~L.} \& \au{Isaacs, J.~D.}} \yr{1977}
  \at{The motion of a small sphere in a rotating velocity field: a possible
  mechanism for suspending particles in turbulence}.  \jt{Journal of
  Geophysical Research}  \bvol{82}~(15),  \pg{2096--2100}.

\bibitem[Toschi \& Bodenschatz(2009)]{Toschi2009}
{\sc \au{Toschi, F.} \& \au{Bodenschatz, E.}} \yr{2009}  \at{Lagrangian
  properties of particles in turbulence}.  \jt{Annual Review of Fluid
  Mechanics}  \bvol{41},  \pg{375--404}.

\bibitem[Vincent \& Meneguzzi(1994)]{vinmen}
{\sc \au{Vincent, A.} \& \au{Meneguzzi, M.}} \yr{1994}  \at{The dynamics of
  vorticity tubes in homogeneous turbulence}.  \jt{Journal of Fluid Mechanics}
  \bvol{258},  \pg{245--254}.

\bibitem[Wang \& Maxey(1993)]{Wang1993}
{\sc \au{Wang, L.~P.} \& \au{Maxey, M.~R.}} \yr{1993}  \at{Settling velocity
  and concentration distribution of heavy particles in homogeneous isotropic
  turbulence}.  \jt{Journal of Fluid Mechanics}  \bvol{256},  \pg{27--68}.

\bibitem[Zhan {\em et~al.\/}(2014)Zhan, Sardina, Lushi \& Brandt]{cj}
{\sc \au{Zhan, C.}, \au{Sardina, G.}, \au{Lushi, E.} \& \au{Brandt, L.}}
  \yr{2014}  \at{Accumulation of motile elongated micro-organisms in
  turbulence}.  \jt{Journal of Fluid Mechanics}  \bvol{739},  \pg{22--36}.

\end{thebibliography}


\begin{thebibliography}{38}
\expandafter\ifx\csname natexlab\endcsname\relax\def\natexlab#1{#1}\fi
\def\au#1{#1} \def\ed#1{#1} \def\yr#1{#1}\def\at#1{#1}\def\jt#1{\textit{#1}}
  \def\bt#1{#1}\def\bvol#1{\textbf{#1}} \def\vol#1{#1} \def\pg#1{#1}
  \def\publ#1{#1}\def\arxiv#1{#1}\def\org#1{#1}\def\st#1{\textit{#1}}

\bibitem[Barton {\em et~al.\/}(2014)Barton, Ward, Williams \&
  Follows]{Barton2014}
{\sc \au{Barton, A.~D.}, \au{Ward, B.~A.}, \au{Williams, R.~G.} \& \au{Follows,
  M.~J.}} \yr{2014}  \at{The impact of fine-scale turbulence on phytoplankton
  community structure}.  \jt{Limnology and Oceanography: Fluids and
  Environments}  \bvol{4}~(1),  \pg{34--49}.

\bibitem[Bec {\em et~al.\/}(2014)Bec, Homann \& Ray]{Bec2014}
{\sc \au{Bec, J.}, \au{Homann, H.} \& \au{Ray, S.~S.}} \yr{2014}
  \at{Gravity-driven enhancement of heavy particle clustering in turbulent
  flow}.  \jt{Physical review letters}  \bvol{112}~(18),  \pg{184501}.

\bibitem[Brenner(1964)]{brenner}
{\sc \au{Brenner, H.}} \yr{1964}  \at{The stokes resistance of an arbitrary
  particleÑiv arbitrary fields of flow}.  \jt{Chemical Engineering Science}
  \bvol{19}~(10),  \pg{703--727}.

\bibitem[Byron {\em et~al.\/}(2015)Byron, Einarsson, Gustavsson, Voth, Mehlig
  \& Variano]{Byron2015}
{\sc \au{Byron, M.}, \au{Einarsson, J.}, \au{Gustavsson, K.}, \au{Voth, G.},
  \au{Mehlig, B.} \& \au{Variano, E.}} \yr{2015}  \at{Shape-dependence of
  particle rotation in isotropic turbulence}.  \jt{Physics of Fluids
  (1994-present)}  \bvol{27}~(3),  \pg{035101}.

\bibitem[Dahlkild(2011)]{Dahlkild2011}
{\sc \au{Dahlkild, A.~A.}} \yr{2011}  \at{Finite wavelength selection for the
  linear instability of a suspension of settling spheroids}.  \jt{Journal of
  Fluid Mechanics}  \bvol{689},  \pg{183--202}.

\bibitem[Daitche(2015)]{Daitche2015}
{\sc \au{Daitche, A.}} \yr{2015}  \at{On the role of the history force for
  inertial particles in turbulence}.  \jt{Journal of Fluid Mechanics}
  \bvol{782},  \pg{567--593}.

\bibitem[Durham {\em et~al.\/}(2013)Durham, Climent, Barry, De~Lillo, Boffetta,
  Cencini \& Stocker]{Durham2013}
{\sc \au{Durham, W.~M.}, \au{Climent, E.}, \au{Barry, M.}, \au{De~Lillo, F.},
  \au{Boffetta, G.}, \au{Cencini, M.} \& \au{Stocker, R.}} \yr{2013}
  \at{Turbulence drives microscale patches of motile phytoplankton}.
  \jt{Nature communications}  \bvol{4}.

\bibitem[Durham {\em et~al.\/}(2011)Durham, Climent \& Stocker]{Durham2011}
{\sc \au{Durham, W.~M.}, \au{Climent, E.} \& \au{Stocker, R.}} \yr{2011}
  \at{Gyrotaxis in a steady vortical flow}.  \jt{Physical Review Letters}
  \bvol{106}~(23),  \pg{238102}.

\bibitem[Eppley {\em et~al.\/}(1967)Eppley, Holmes \& Strickland]{Eppley1967}
{\sc \au{Eppley, R.~W.}, \au{Holmes, R.~W.} \& \au{Strickland, J. D.~H.}}
  \yr{1967}  \at{Sinking rates of marine phytoplankton measured with a
  fluorometer}.  \jt{Journal of Experimental Marine Biology and Ecology}
  \bvol{1}~(2),  \pg{191--208}.

\bibitem[Fornari {\em et~al.\/}(2016)Fornari, Picano \& Brandt]{Fornari2016}
{\sc \au{Fornari, W.}, \au{Picano, F.} \& \au{Brandt, L.}} \yr{2016}
  \at{Sedimentation of finite-size spheres in quiescent and turbulent
  environments}.  \jt{Journal of Fluid Mechanics}  \bvol{788},  \pg{640--669}.

\bibitem[Gallily \& Cohen(1979)]{Gallily1979}
{\sc \au{Gallily, I.} \& \au{Cohen, A.~H.}} \yr{1979}  \at{On the orderly
  nature of the motion of nonspherical aerosol particles. ii. inertial
  collision between a spherical large droplet and an axially symmetrical
  elongated particle}.  \jt{Journal of Colloid and Interface Science}
  \bvol{68}~(2),  \pg{338--356}.

\bibitem[Guazzelli \& Morris(2011)]{Guazzelli2011}
{\sc \au{Guazzelli, E.} \& \au{Morris, J.~F.}} \yr{2011} {\em A physical
  introduction to suspension dynamics\/}, ,  \vol{vol.~45}.  \publ{Cambridge
  University Press}.

\bibitem[Gustavsson {\em et~al.\/}(2016)Gustavsson, Berglund, Jonsson \&
  Mehlig]{Gustavsson2016}
{\sc \au{Gustavsson, K.}, \au{Berglund, F.}, \au{Jonsson, P.~R.} \& \au{Mehlig,
  B.}} \yr{2016}  \at{Preferential sampling and small-scale clustering of
  gyrotactic microswimmers in turbulence}.  \jt{Physical review letters}
  \bvol{116}~(10),  \pg{108104}.

\bibitem[Gustavsson {\em et~al.\/}(2014)Gustavsson, Vajedi \&
  Mehlig]{Gustavsson2014}
{\sc \au{Gustavsson, K.}, \au{Vajedi, S.} \& \au{Mehlig, B.}} \yr{2014}
  \at{Clustering of particles falling in a turbulent flow}.  \jt{Physical
  Review Letters}  \bvol{112}~(21),  \pg{214501}.

\bibitem[Jeffery(1922)]{jeff}
{\sc \au{Jeffery, G.~B.}} \yr{1922} The motion of ellipsoidal particles
  immersed in a viscous fluid.  \bt{In {\em Proceedings of the Royal Society of
  London A: Mathematical, Physical and Engineering Sciences\/}}, ,  \vol{vol.
  102},  \pg{pp. 161--179}. The Royal Society.

\bibitem[Karp-Boss \& Jumars(1998)]{Karp1998}
{\sc \au{Karp-Boss, L.} \& \au{Jumars, P.~A.}} \yr{1998}  \at{Motion of diatom
  chains in steady shear flow}.  \jt{OCEANOGRAPHY}  \bvol{43}~(8).

\bibitem[Lazier \& Mann(1989)]{Lazier1989}
{\sc \au{Lazier, J. R.~N.} \& \au{Mann, K.~H.}} \yr{1989}  \at{Turbulence and
  the diffusive layers around small organisms}.  \jt{Deep Sea Research Part A.
  Oceanographic Research Papers}  \bvol{36}~(11),  \pg{1721--1733}.

\bibitem[Lucci {\em et~al.\/}(2011)Lucci, Ferrante \& Elghobashi]{Lucci2011}
{\sc \au{Lucci, F.}, \au{Ferrante, A.} \& \au{Elghobashi, S.}} \yr{2011}
  \at{Is stokes number an appropriate indicator for turbulence modulation by
  particles of taylor-length-scale size?}  \jt{Physics of Fluids
  (1994-present)}  \bvol{23}~(2),  \pg{025101}.

\bibitem[Marchioli {\em et~al.\/}(2014)Marchioli, Zhao \& Andersson]{zhao}
{\sc \au{Marchioli, C.}, \au{Zhao, L.} \& \au{Andersson, H.~I.}} \yr{2014}
  \at{On the relative rotational motion between rigid fibers and fluid in
  turbulent channel flow}.  \jt{Physics of Fluids (1994-present)}
  \bvol{28}~(1),  \pg{013301}.

\bibitem[Maxey(1987)]{Maxey1987}
{\sc \au{Maxey, M.~R.}} \yr{1987}  \at{The gravitational settling of aerosol
  particles in homogeneous turbulence and random flow fields}.  \jt{Journal of
  Fluid Mechanics}  \bvol{174},  \pg{441--465}.

\bibitem[Murray(1970)]{Murray1970}
{\sc \au{Murray, S.~P.}} \yr{1970}  \at{Settling velocities and vertical
  diffusion of particles in turbulent water}.  \jt{Journal of geophysical
  research}  \bvol{75}~(9),  \pg{1647--1654}.

\bibitem[Nguyen {\em et~al.\/}(2011)Nguyen, Karp-Boss, Jumars \&
  Fauci]{Nguyen2011}
{\sc \au{Nguyen, H.}, \au{Karp-Boss, L.}, \au{Jumars, P.~A} \& \au{Fauci, L.}}
  \yr{2011}  \at{Hydrodynamic effects of spines: A different spin}.
  \jt{Limnology and Oceanography: Fluids and Environments}  \bvol{1}~(1),
  \pg{110--119}.

\bibitem[Ni {\em et~al.\/}(2015)Ni, Kramel, Ouellette \& Voth]{Ni2015}
{\sc \au{Ni, R.}, \au{Kramel, S.}, \au{Ouellette, N.~T} \& \au{Voth, G.~A.}}
  \yr{2015}  \at{Measurements of the coupling between the tumbling of rods and
  the velocity gradient tensor in turbulence}.  \jt{Journal of Fluid Mechanics}
   \bvol{766},  \pg{202--225}.

\bibitem[Nielsen(1993)]{Nielsen1993}
{\sc \au{Nielsen, P.}} \yr{1993}  \at{Turbulence effects on the settling of
  suspended particles}.  \jt{Journal of Sedimentary Research}  \bvol{63}~(5).

\bibitem[Nielsen(2007)]{Nielsen2007}
{\sc \au{Nielsen, P.}} \yr{2007}  \at{Mean and variance of the velocity of
  solid particles in turbulence}.  \bt{In {\em Particle-Laden Flow\/}},
  \pg{pp. 385--391}.  \publ{Springer}.

\bibitem[Olivieri(2013)]{Olivieri2013}
{\sc \au{Olivieri, S.}} \yr{2013} Analysis of the forces acting on particles in
  homogeneous isotropic turbulence.

\bibitem[Olivieri {\em et~al.\/}(2014)Olivieri, Picano, Sardina, Ludicone \&
  Brandt]{Olivieri2014}
{\sc \au{Olivieri, S.}, \au{Picano, F.}, \au{Sardina, G.}, \au{Ludicone, D.} \&
  \au{Brandt, L.}} \yr{2014}  \at{The effect of the basset history force on
  particle clustering in homogeneous and isotropic turbulence}.  \jt{Physics of
  Fluids (1994-present)}  \bvol{26}~(4),  \pg{041704}.

\bibitem[Parsa \& Voth(2014)]{Parsa2014}
{\sc \au{Parsa, S.} \& \au{Voth, G.~A.}} \yr{2014}  \at{Inertial range scaling
  in rotations of long rods in turbulence}.  \jt{Physical review letters}
  \bvol{112}~(2),  \pg{024501}.

\bibitem[Rogallo(1981)]{rogallo}
{\sc \au{Rogallo, R.~S.}} \yr{1981}  \at{Numerical experiments in homogeneous
  turbulence}.  \jt{NASA STI/Recon Technical Report N}  \bvol{81},  \pg{31508}.

\bibitem[Ruiz {\em et~al.\/}(2004)Ruiz, Mac{\'\i}as \& Peters]{Ruiz2004}
{\sc \au{Ruiz, J.}, \au{Mac{\'\i}as, D.} \& \au{Peters, F.}} \yr{2004}
  \at{Turbulence increases the average settling velocity of phytoplankton
  cells}.  \jt{Proceedings of the National academy of Sciences of the United
  States of America}  \bvol{101}~(51),  \pg{17720--17724}.

\bibitem[Sardina {\em et~al.\/}(2015)Sardina, Picano, Brandt \&
  Caballero]{prlcloud}
{\sc \au{Sardina, G.}, \au{Picano, F.}, \au{Brandt, L.} \& \au{Caballero, R.}}
  \yr{2015}  \at{Continuous growth of droplet size variance due to condensation
  in turbulent clouds}.  \jt{Physical review letters}  \bvol{115}~(18),
  \pg{184501}.

\bibitem[Smayda(1970)]{Smayda1970}
{\sc \au{Smayda, T.~J.}} \yr{1970}  \at{The suspension and sinking of
  phytoplankton in the sea} .

\bibitem[Smyth \& Moum(2000)]{Smyth2000}
{\sc \au{Smyth, W.~D.} \& \au{Moum, J.~N.}} \yr{2000}  \bt{Ocean turbulence}.
  {\em Tech. Rep.\/}.  \org{Technical report, College of Oceanic and
  Atmospheric Sciences, Oregon State University}.

\bibitem[Tooby {\em et~al.\/}(1977)Tooby, Wick \& Isaacs]{Tooby1977}
{\sc \au{Tooby, P.~F.}, \au{Wick, G.~L.} \& \au{Isaacs, J.~D.}} \yr{1977}
  \at{The motion of a small sphere in a rotating velocity field: a possible
  mechanism for suspending particles in turbulence}.  \jt{Journal of
  Geophysical Research}  \bvol{82}~(15),  \pg{2096--2100}.

\bibitem[Toschi \& Bodenschatz(2009)]{Toschi2009}
{\sc \au{Toschi, F.} \& \au{Bodenschatz, E.}} \yr{2009}  \at{Lagrangian
  properties of particles in turbulence}.  \jt{Annual Review of Fluid
  Mechanics}  \bvol{41},  \pg{375--404}.

\bibitem[Vincent \& Meneguzzi(1994)]{vinmen}
{\sc \au{Vincent, A.} \& \au{Meneguzzi, M.}} \yr{1994}  \at{The dynamics of
  vorticity tubes in homogeneous turbulence}.  \jt{Journal of Fluid Mechanics}
  \bvol{258},  \pg{245--254}.

\bibitem[Wang \& Maxey(1993)]{Wang1993}
{\sc \au{Wang, L.~P.} \& \au{Maxey, M.~R.}} \yr{1993}  \at{Settling velocity
  and concentration distribution of heavy particles in homogeneous isotropic
  turbulence}.  \jt{Journal of Fluid Mechanics}  \bvol{256},  \pg{27--68}.

\bibitem[Zhan {\em et~al.\/}(2014)Zhan, Sardina, Lushi \& Brandt]{cj}
{\sc \au{Zhan, C.}, \au{Sardina, G.}, \au{Lushi, E.} \& \au{Brandt, L.}}
  \yr{2014}  \at{Accumulation of motile elongated micro-organisms in
  turbulence}.  \jt{Journal of Fluid Mechanics}  \bvol{739},  \pg{22--36}.

\end{thebibliography}


\begin{thebibliography}{14}
\expandafter\ifx\csname natexlab\endcsname\relax\def\natexlab#1{#1}\fi

\bibitem[Batchelor(1971)]{Batchelor59}
{\sc Batchelor, G.~K.} 1971 Small-scale variation of convected quantities like
  temperature in turbulent fluid. part 1. general discussion and the case of
  small conductivity. {\em J.~Fluid Mech.\/} {\bf 5}, 113--133.

\bibitem[Brownell \& Su(2004)]{Brownell04}
{\sc Brownell, C.~J. \& Su, L.~K.} 2004 Planar measurements of differential
  diffusion in turbulent jets. {\em AIAA Paper 2004-2335\/}.

\bibitem[Brownell \& Su(2007)]{Brownell07}
{\sc Brownell, C.~J. \& Su, L.~K.} 2007 Scale relations and spatial spectra in
  a differentially diffusing jet. {\em AIAA Paper 2007-1314\/}.

\bibitem[Dennis(1985)]{Dennis85}
{\sc Dennis, S. C.~R.} 1985 {Compact explicit finite difference approximations
  to the Navier--Stokes equation}. In {\em Ninth Intl Conf. on Numerical
  Methods in Fluid Dynamics\/} (ed. Soubbaramayer \& J.~P. Boujot), {\em
  Lecture Notes in Physics\/}, vol. 218, pp. 23--51. Springer.

\bibitem[Hwang \& Tuck(1970)]{Hwang70}
{\sc Hwang, L.-S. \& Tuck, E.~O.} 1970 On the oscillations of harbours of
  arbitrary shape. {\em J.~Fluid Mech.\/} {\bf 42}, 447--464.

\bibitem[Koch(1983)]{Koch83}
{\sc Koch, W.} 1983 Resonant acoustic frequencies of flat plate cascades. {\em
  J.~Sound Vib.\/} {\bf 88}, 233--242.

\bibitem[Lee(1971)]{Lee71}
{\sc Lee, J.-J.} 1971 Wave-induced oscillations in harbours of arbitrary
  geometry. {\em J.~Fluid Mech.\/} {\bf 45}, 375--394.

\bibitem[Linton \& Evans(1992)]{Linton92}
{\sc Linton, C.~M. \& Evans, D.~V.} 1992 The radiation and scattering of
  surface waves by a vertical circular cylinder in a channel. {\em Phil.\
  Trans.\ R. Soc.\ Lond.\/} {\bf 338}, 325--357.

\bibitem[Martin(1980)]{Martin80}
{\sc Martin, P.~A.} 1980 On the null-field equations for the exterior problems
  of acoustics. {\em Q.~J. Mech.\ Appl.\ Maths\/} {\bf 33}, 385--396.

\bibitem[Miller(1991)]{Miller91}
{\sc Miller, P.~L.} 1991 Mixing in high schmidt number turbulent jets. PhD
  thesis, California Institute of Technology.

\bibitem[Rogallo(1981)]{Rogallo81}
{\sc Rogallo, R.~S.} 1981 Numerical experiments in homogeneous turbulence. {\em
  Tech. Rep.\/} 81835. NASA Tech.\ Mem.

\bibitem[Ursell(1950)]{Ursell50}
{\sc Ursell, F.} 1950 Surface waves on deep water in the presence of a
  submerged cylinder i. {\em Proc.\ Camb.\ Phil.\ Soc.\/} {\bf 46}, 141--152.

\bibitem[{van Wijngaarden}(1968)]{Wijngaarden68}
{\sc {van Wijngaarden}, L.} 1968 On the oscillations near and at resonance in
  open pipes. {\em J.~Engng Maths\/} {\bf 2}, 225--240.

\bibitem[Worster(1992)]{Worster92}
{\sc Worster, M.~G.} 1992 {The dynamics of mushy layers}. In {\em In
  Interactive dynamics of convection and solidification\/} (ed. S.~H. Davis,
  H.~E. Huppert, W.~Muller \& M.~G. Worster), pp. 113--138. Kluwer.

\end{thebibliography}

\end{document}